\documentclass[11pt,a4paper,oneside]{article} 
\usepackage[T1]{fontenc} 
\usepackage[pdftex]{graphicx}
\usepackage[hmargin=3cm,vmargin={2.5cm,3cm}]{geometry}
\usepackage{amsmath,amsfonts,amssymb}
\usepackage[english]{babel}
\usepackage[pdftex,hyperfigures,breaklinks,colorlinks,citecolor=blue,linkcolor=red]{hyperref}
\usepackage{authblk}

\def\be{\begin{equation}}
\def\ee{\end{equation}}
\def\bearr{\begin{eqnarray}}
\def\eearr{\end{eqnarray}}
\def\bc{\begin{center}}
\def\ec{\end{center}}

\def\up{\textsuperscript}
\def\<{\bigl\langle}
\def\>{\bigr\rangle}

\def\ll{\langle}
\def\rr{\rangle}

\def\Q{\mathrm{Q}}

\def\s{\sigma}

\title{Perturbative large deviation analysis of non-equilibrium dynamics}

\author[1]{Gino Del Ferraro\footnote{Email address: gino@kth.se}}
\author[1,2]{Erik Aurell \footnote{Email address: eaurell@kth.se, erik.aurell@aalto.fi}}
\affil[1]{\small Department of Computational Biology, KTH-Royal
Institute of Technology, SE-100 44 Stockholm, Sweden} 
\affil[2]{ Department of
Information and Computer Science, Aalto University, FIN-00076 Aalto, Finland}

\begin{document}
\maketitle

\begin{abstract}
Macroscopic fluctuation theory has shown that a wide class of non-equilibrium
stochastic dynamical systems obey a large deviation principle, but except for
a few one-dimensional examples these large deviation 
principles are in general not known in closed form. We consider 
the problem of constructing successive approximations to an (unknown)
large deviation functional and show
that the 
non-equilibrium probability distribution the takes a Gibbs-Boltzmann form
with a set of auxiliary (non-physical) energy functions.
The expectation values of these auxiliary energy functions and their conjugate quantities satisfy a closed 
system of equations which can imply a considerable reduction of dimensionality of the dynamics.
We show that the accuracy of the approximations can be tested self-consistently
without solving the full non-equilibrium equations.
We test the general procedure on the simple model 
problem of a relaxing 1D Ising chain.
\end{abstract}

\section{Introduction}

Non-equilibrium dynamics is central to diverse interdisciplinary applications of statistical
physics to fields such as neuroscience~\cite{HertzKroghPalmer1991,Schneidman06,RoudiTyrchaHertz09PRE},
gene regulatory systems~\cite{Huang2009},
socio-economic systems~\cite{MategnaStanley,Maslov2000,BouchaudPotters} 
large-scale combinatorial optimization 
problems~\cite{BoettcherPercus2001,BarthelHartmannWeigt2003,MonassonSemerjian2003,Alava2008},
and models of evolution in biology and other fields~\cite{BlytheMcKane2007}.
With the increasing amount of data available, on all kinds of large systems, systematic
methods to analyze large non-equilibrium models have therefore come into focus.
If one is interested in the dominant features of
a system comprised of a large number $N$ of variables, and if a probability
distribution over the system 
obeys a large deviation principle
$P({s})\propto e^{-NV(s)}$, where $V$ is a large deviation functional~\cite{Varadhan1984},
then a general approach is to determine and analyze $V$.
Important progress has been made in the macroscopic fluctuation theory~\cite{Bertini2002},
where been shown that for spatially extended systems of the diffusion
type there is a deviation functional which obeys a variational principle hich can be solved
exactly for certain one-dimensional models~\cite{Bertini2007,Derrida2007}. Recently this
variational principle was 
also solved for a weakly mean-field coupled non-equilibrium system~\cite{Gawedzki2013}.

In this paper we consider the problem if an (unknown) large deviation principle can be approximated
in a perturbation theory.   
At each level of approximation the non-equilibrium probability distribution is then 
described as a Gibbs measure with an auxiliary (non-physical) energy function,
as discussed in Appendix~A of~\cite{EyinkLebowitzSpohn1996}. 
The expectation values of the terms in this  
auxiliary energy function and their conjugate quantities satisfy a closed set
system of equations, potentially, as we will discuss,
a considerable reduction of dimensionality of the dynamics.
This fact was, to our knowledge, and using different arguments, first used as 
a computational scheme in (fully-connected) mean field models of spin glasses~\cite{DRTinf1,DRTinf2}
later extended to diluted systems on random graphs~\cite{Hatchett,Mozeika1,Mozeika2}.
Here we consider this approach from the viewpoint of large deviations with a focus
on how the accuracy of the approximation can be assessed internally without resorting to
simulations of the full dynamics. 
 
The paper is organized as follows: in Section~\ref{sec:LD} we introduce the general approach,
and in Section~\ref{sec:perturbation} we show how it can in principle be used to set up a perturbation
theory. In Sections~\ref{sec:2parameters} and~\ref{sec:JSF} we analyze the model problem
of a relaxing Ising chain, first on the level of approximation of a ``two-parameter theory'', and then
on the level of the ``joint spin-field theory'' a higher-order approximation
developed in~\cite{DRTinf1,DRTinf2,Hatchett,Mozeika1,Mozeika2}.
We here show that the accuracy of each level of approximation can be assessed internally,
where the largest inaccuracies of the ``two-parameter theory'' are taken care of in the
``joint spin-field theory'', but where higher-order inaccuracies would need a higher-order approximation.
In Section~\ref{sec:discussion} we summarize and discuss our results.
Four appendices contain additional material: in Appendix~\ref{appendix1}
we give the details of our derivation of the joint spin-field theory by a graph inflation
procedure combined with ordinary Belief Propagation, while in Appendix~\ref{appendix2} we
do the same using the approach developed in~\cite{DRTinf1,DRTinf2,Hatchett,Mozeika1,Mozeika2}. 
The resulting equations are in both cases rather complicated, and we therefore
show separately in Appendix~\ref{appendix3} that they indeed lead to the same computational scheme.
In the last Appendix~\ref{appendix4} we show that our approach
can be given a geometric interpretation as a projection on an $e$-flat hierarchy of probability
distributions, as defined in~\cite{Amari2001},
at each time step of the dynamics. We also show that the internal test for accuracy which we
develop has the counter-party in the dual concept of projection on $m$-flat hierarchies.  

\section{Large deviations and the dimensional reduction of dynamics}
\label{sec:LD}
For definiteness we will consider a continuous-time Markov process on $N$ Boolean
variables (spins) described by a master equation
\begin{equation}\label{eq:master-equation}
 \partial_tP({s}) = \sum_{j=1}^N w_j(F_j {s})P(F_j{s})-  w_j({s})P({s})
\end{equation}
where  $w_j({s})$ is the flip rate of spin $s_j$ and $F_j$ is the flip operator
\textit{i.e.} $F_j{s}=F_j(s_1,\ldots,s_j,\ldots,s_N)=(s_1,\ldots,-s_j,\ldots,s_N)$.
We will assume 
that the probability distribution takes a large deviation form $P({s})\propto \exp\left(-NV({s}\right))$ 
and that the large deviation function $V$ depends on $L$ 
intrinsic quantities (homogeneous local functions) $V=V\left(o_1({s}),\ldots,o_L({s})\right)$.
When the variations are (relatively) small the large deviation function can be linearized such that 
$V=\hbox{Const.}+\beta_1o_1({s})+\ldots+\beta_Lo_L({s})$ where now
$O_1({s})=No_1({s}),\ldots,O_L({s})=No_L({s})$ are auxiliary (non-physical) energy terms, 
$\beta_1,\ldots\beta_L$ the corresponding conjugate quantities (generalized temperatures), and
the (non-equilibrium) probability distribution is approximated by an auxiliary Gibbs measure
\begin{equation}
P^{\hbox{\small aux}}({s}) = \exp\left(-\beta_1 O_1({s})\ldots-\beta_L O_L({s})-F\right)
\label{eq:auxiliary-Gibbs}
\end{equation}
where $F$ is the normalization (topological pressure).
We define the expectation values of the intrinsic quantities as 
$\mu_1=<o_1>$,$\ldots$, $\mu_L=<o_L>$
and assume that $P$ and $P^{\hbox{\small aux}}$ are close enough that the
the expectations can be taken
with respect to either with the same results up to terms of order $1/N$. 
In Appendix~\ref{appendix4} we relate this approximation to the projection on $e$-flat hierarchies of probability distributions
as has been defined in Information Geometry~\cite{Amari2001}. 
   
When the dynamics under (\ref{eq:master-equation}) changes the probability from $P$ to $P'=P+\delta P$ 
the $\mu$'s change as
\begin{equation}
\frac{d^{(T)}\mu_l}{dt}=\frac{1}{N}\sum_{j=1}^N \left<w_j\Delta_{O_l\to j}\right> 
\label{eq:mu-beta-dynamics}   
\end{equation}
where $\Delta_{O_l\to j}({s})= O_l(F_j{s})-O_l({s})$, and where the superscript $T$
indicates that this is the change following the true dynamics.
We can also consider the same expectation values with respect to the measure (\ref{eq:auxiliary-Gibbs})
before and after an infinitesimal change of the generalized temperatures which gives
\begin{equation}
\frac{d^{(M)}\mu_l}{dt}=-\sum_{n} C_{ln}\dot{\beta_n}\qquad   C_{ln}=\frac{1}{N}\left[<O_l(s)O_n(s)>-<O_l(s)><O_n(s)>\right]
\label{eq:mu-beta-dynamics-M}   
\end{equation}
where the superscript $M$ indicates that the change follows the model.
Assuming for simplicity that $C$, the covariance matrix of the energy terms, has full rank, and
setting (\ref{eq:mu-beta-dynamics}) equal to (\ref{eq:mu-beta-dynamics-M}) for the expectation
values $\mu_l$ of all auxiliary energy terms in (\ref{eq:auxiliary-Gibbs}), we have
an equation for the rate of change of the generalized temperatures:
\begin{equation}
\dot{\beta_l}=-\sum_{n} C^{-1}_{ln} \left[\frac{1}{N}\sum_{j=1}^N \left<w_j\Delta_{O_n\to j}\right>\right] 
\label{eq:beta-dynamics}   
\end{equation}
where $C^{-1}$ is the inverse of $C$. 

The two equations (\ref{eq:mu-beta-dynamics})  and (\ref{eq:mu-beta-dynamics-M}) also
make sense for an observable \textit{not} included in the model. If $Q(s)$ is
such an observable and $\mu_q$ is the expectation value of $q(s)=Q(s)/N$, then
\begin{equation}
\frac{d^{(T)}\mu_q}{dt}-\frac{d^{(M)}\mu_q}{dt}
=\frac{1}{N}\sum_{j=1}^N \left<w_j\Delta_{Q\to j}\right> 
-\sum_{nl} C_{Ql}C^{-1}_{ln}\frac{1}{N}\sum_{j=1}^N \left<w_j\Delta_{O_n\to j}\right>  
\label{eq:error-Q}   
\end{equation}
which is zero if $Q$ is one of the $O_n$'s, but otherwise does not have to vanish.
This difference is hence an internal quality check which can be used to estimate
if the distribution $P^{\hbox{\small aux}}$ is (locally) a good approximation to $P$, in the direction
of observable $Q$. 
In Appendix~\ref{appendix4} we relate this concept of approximation quality in the direction of
an observable to the projection on $m$-flat hierarchies of probability distributions
(dual to $e$-flat hierarchies) as was introduced in Information Geometry~\cite{Amari2001}. 

The computation of the time derivatives of the expectation values
(\ref{eq:mu-beta-dynamics}) and the self-consistency tests (\ref{eq:error-Q})
both reduce to computing  marginal probabilities over subsets of variables with respect to (\ref{eq:auxiliary-Gibbs}).
This is in general (if done exactly) of exponential complexity in systems size~\cite{MMbook}, 
and the reduction therefore does not in itself simplify the problem to understand the dynamics of (\ref{eq:master-equation}).
However, if and when these marginals can be computed accurately
by mean field methods, or Belief Propagation~\cite{MMbook}, or generalizations thereof~\cite{YedidiaFreemanWeiss2003},
then the dimensionality is reduced from $2^N-1$ to polynomial in $N$ while remaining efficiently computable.
This is the reduction of dimensionality of dynamics which is the topic of this paper.

We remark that though (\ref{eq:beta-dynamics}) is in principle exact it is not practically useful, as 
the covariance matrix of the energy terms is typically cumbersome to compute.
Instead one may use the definition $\mu_l=<o_l>$ and solve the inverse problem of computing
the $\beta$'s from the $\mu$'s before and after changing the $\mu$'s, using (\ref{eq:mu-beta-dynamics}).
In the following we will assume that this inverse problem can be solved once for the
initial state, either by a brute force approach or, for instance, by requiring that initially the spins are independent such that most of
the generalized temperatures in (\ref{eq:auxiliary-Gibbs}) are initially zero. Further changes in the $\beta$'s
can then found incrementally \textit{e.g.} by using a Newton-Raphson routine.

\section{Perturbative large deviations for non-equilibrium dynamics}
\label{sec:perturbation}
\begin{figure}[!htb]
\begin{center}
\includegraphics[width=6cm]{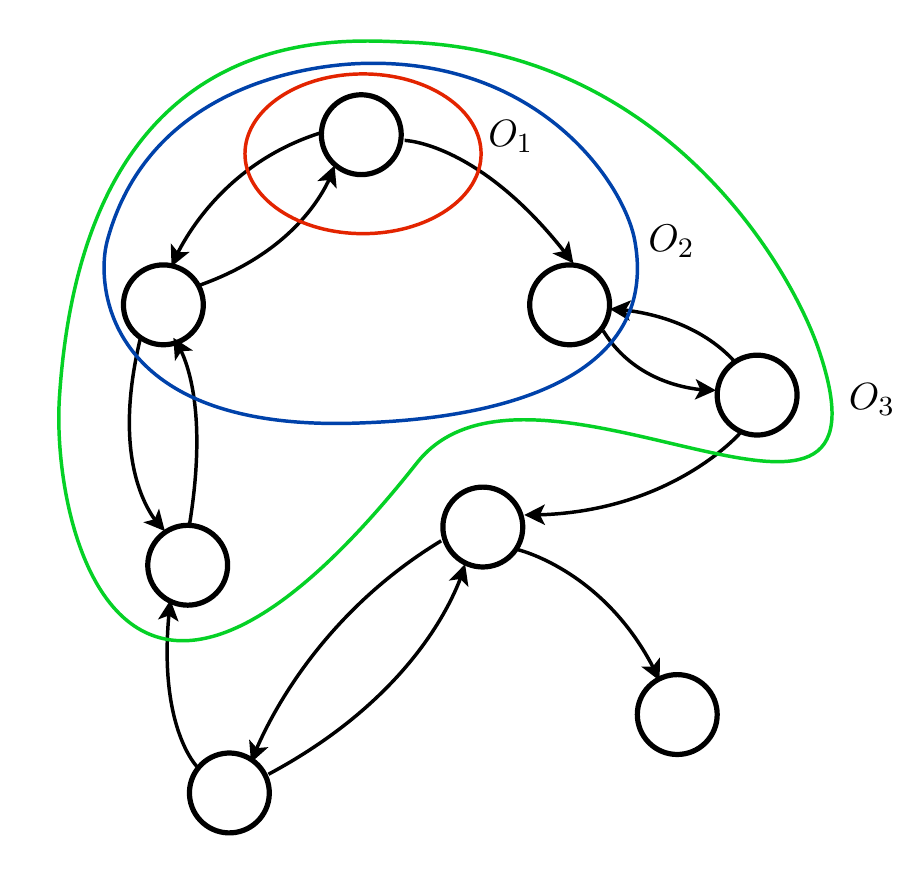}  
\caption{\footnotesize A dependency graph $G$ with neighborhoods containing respectively a single spin ($O_1$),
a single spin and its parents in the dependency graph ($O_2$), and a single spin and its parents and grandparents in the
dependency graph ($O_3$) . It is suggested that putative auxiliary energy terms 
in (\protect\ref{eq:auxiliary-Gibbs}) can be classified as containing
spins in sets of of the type $O_1$, $O_2$, $O_3$ etc. (see main text).}
\label{fig:network}
\end{center}
\end{figure}

We consider the evaluation of the
time derivatives $\dot{\mu_1},\ldots,\dot{\mu_L}$, which,
as already observed, amounts to determining marginal probabilities of the auxiliary Gibbs measure.
The auxiliary measure introduce in (\ref{eq:auxiliary-Gibbs}) can be described by a factor graph $F^{\hbox{\small aux}}$~\cite{MMbook}
and, as is well known, marginal probabilities can be efficiently computed 
if the factor graph is locally tree-like~\cite{MMbook}.  
The master equation (\ref{eq:master-equation}) on the other hand defines a (directed) dependency graph $G$ where 
the vertices stand for spins and where there is a link from spin $i$ to spin $j$ if the rate $w_j$ depends on spin $s_i$. 
Putative interactions in  $F^{\hbox{\small aux}}$ can then be partitioned as to how distant are its terms in $G$.
We will posit that a set 
$O_1(s)$ contains all (auxiliary, non-physical) interactions depending on one spin only, 
a set $O_2(s)$ contains all (auxiliary, non-physical) interactions between a spin $s_j$ 
and the set of spins $s_i$ such that the (physical flipping) rate $w_j$ depends on $s_i$, and so on.
The concept is explained in Fig.~\ref{fig:network}.
It is reasonable to expect that the more terms $O_1(s),O_2(s),\ldots$ are included in the auxiliary probability distribution (\ref{eq:auxiliary-Gibbs}), 
the more accurate may be the approximations to a full probability distribution $P$. 
This is then a possible basis of a perturbation expansion, which however, as we will see,
comes at the price of somewhat quickly increasing complexity. 

\begin{figure}[!htb]
\begin{center}
\includegraphics[width=6cm]{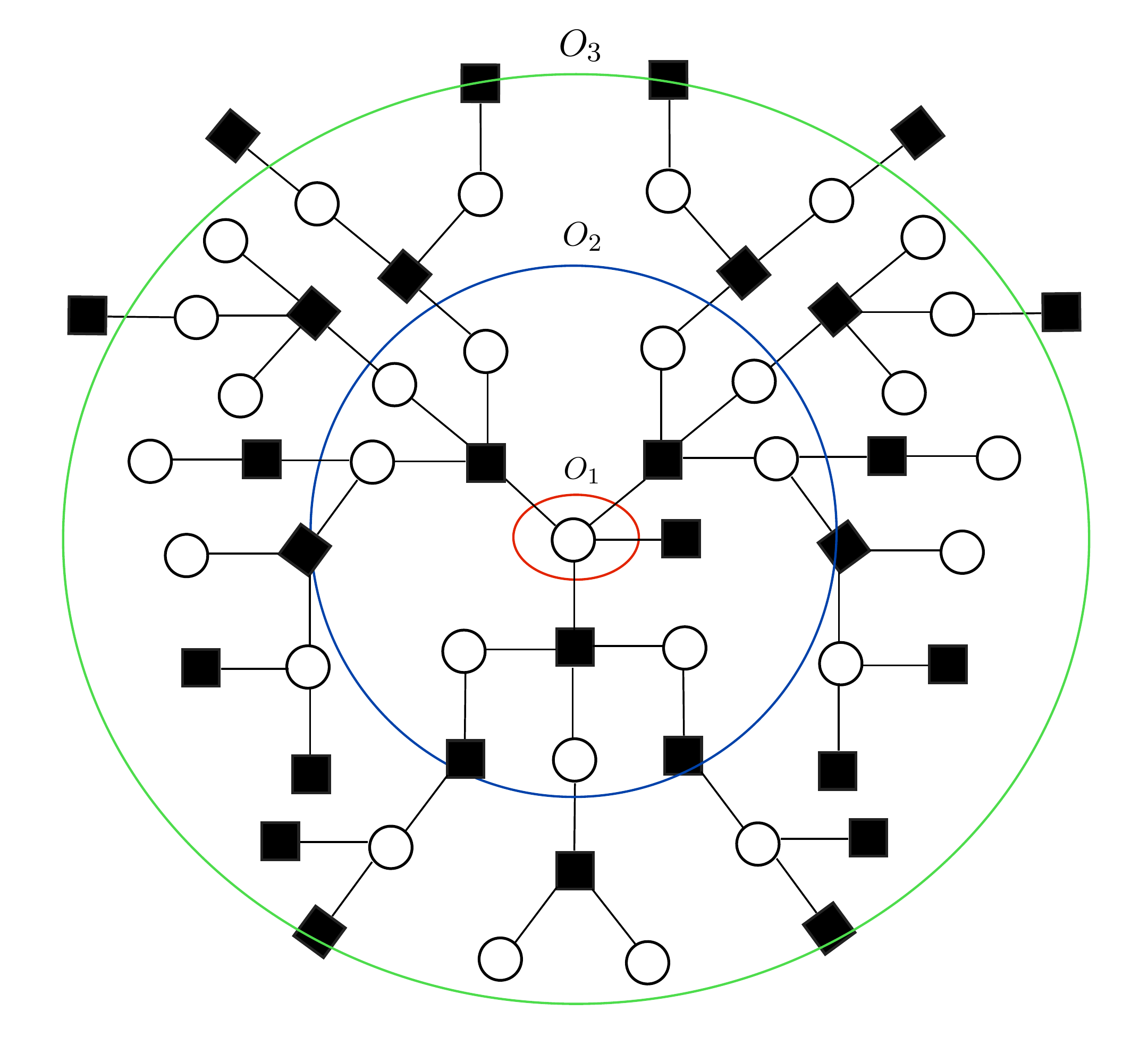}  
\caption{\footnotesize A factor graph $F$ describing an energy function governing a dynamics.
This graph relates to the more general formulation in Fig.~(\protect\ref{fig:network}) in that
two nodes in $G$ will be (symmetrically) connected if they are at distance two in $F$, separated by an
energy term which depends on both variables. Neighborhoods $O_1$, $O_2$ and $O_3$ are defined
analogously to Fig.~(\protect\ref{fig:network}).}
\label{fig:network-F}
\end{center}
\end{figure}

For simplicity we will from now assume that the dynamics in the master equation follows from an energy function such that 
the rates $w_j$ depend on the values of the local energy terms and on how these change if spin $\sigma_j$ would be flipped. 
This setting includes systems 
obeying detailed balance (and which relax to equilibrium), but also diffusive systems driven by boundary 
terms or bulk drift~\cite{EyinkLebowitzSpohn1996,Bertini2002,Bertini2007,Derrida2007} and versions of
focused local search on large random graphs~\cite{Alava2008,Lemoy2013}.
The (physical) energy function is then also described by a factor graph $F$, and the computational properties
of an auxiliary Gibbs measure can be discussed in how terms in how $F^{\hbox{\small aux}}$ 
relate to $F$. The concept is explained in Fig.~\ref{fig:network-F}. 

A theory containing only terms in $O_1$ is one where the spins are assumed independent,
and this factorized probability distribution is used to compute the expectation values
in (\ref{eq:mu-beta-dynamics}). This can be contrasted to dynamical mean-field theory
which starts from the equivalent of (\ref{eq:mu-beta-dynamics}) and the expectation
value computed with respect to the full probability distribution, which is then expanded in a perturbation
series in interaction strength~\cite{Kappen2000mean,Tanaka2000,RoudiHertz11JSAT}.
In dynamic mean field theory there is the notion that one seeks the factorized probability
distribution which has smallest Kullback-Leibler divergence from the full distribution, 
without assuming that the factorized probability distribution and the full distribution are actually close.
In the approach developed here we on the other hand assume that $P$ is always close to $P^{\hbox{aux}}$, 
which for a theory containing terms only in $O_1$ means a factorized probability distribution,
and use (\ref{eq:mu-beta-dynamics-M}) and (\ref{eq:beta-dynamics}) to enforce this condition at each time step.   

Theories containing only terms in $O_2$ fall naturally into two categories.
In the first we restrict the allowed interactions in $F^{\hbox{\small aux}}$ 
to be as in $F$ and $O_1(s)$ (meaning external fields, in the case these are not included in $F$).
$F^{\hbox{\small aux}}$ will then have the same topology as $F$ and for Ising pair-wise interactions (including the 
example studied below) this means a theory depending on magnetizations (one-spin marginals) and energies (two-spin marginals,
where the two spins are at distance two in $F$). Such a theory, while still not trivial, 
has the advantage that marginal probabilities can be computed in the same manner for
$F^{\hbox{\small aux}}$ as for $F$, typically by the cavity method.
For the example of a relaxing Ising spin chain we develop the theory on this level
of approximation in Section \ref{sec:2parameters}. 
We note that if we would have interactions in $F$ among three and
more spins then
we can have many more terms in the auxiliary Gibbs measure than magnetization and
energy, on this level of approximation.
One example would be to include in the description
of a 3-spin interacting system both the physical energy on three spins (say $s_i s_j s_k$), and also
other terms depending on the same spins but in a different manner (say $s_i s_j + s_j s_k + s_k s_i$).
This category of computationally comparatively simple theories in  $O_2$ is therefore in general larger than
theories based only on magnetization and energy.

The second category of theories containing only terms in $O_2$ are the
rest, where the factor graph $F^{\hbox{\small aux}}$ does not have the same
topology as $F$, and typically is not locally tree-like.
If the rate $w_j$ depends on a set of local energy terms $\epsilon_a(s^a)$ where $j$ and $a$
are linked in the factor graph $F$ then the set $O_2(s)$ contains, in general, all interactions depending on
$s_j$ and on any of the other spins in the sets $s^a$, but not on any other spins.
This set is larger than the interactions included in $F$ (the first category)
because terms depending on a spin $s_j$ and spins in at least two
different energy terms $a$ and $b$, both linked to $j$, are included.
Theories in this category are necessarily more complicated since the
marginal probabilities cannot be computed in the same way for $F^{\hbox{\small aux}}$ as for $F$.
We will below in Section~\ref{sec:JSF} develop one such approximation
for the relaxing Ising spin chain which we refer to as the ``joint spin-field theory'',
following the earlier literature~\cite{DRTinf2,Hatchett,Mozeika1,Mozeika2}. 
We will show how the abundance of short loops can then be handled by a graph inflation
technique such that the expectation values can be computed by ordinary BP on 
an auxiliary locally tree-like graph (the expanded graph). Details are
and comparisons to the approach taken in~\cite{DRTinf2,Hatchett,Mozeika1,Mozeika2}
are given in Appendices~\ref{appendix1}-\ref{appendix3}.

Theories containing only terms in $O_3$ (and higher orders) will not be considered in detail
in this paper. It is however clear that they pose similar problems as the category of
general theories containing only terms in $O_2$, \textit{i.e.} that the auxiliary factor
graph  $F^{\hbox{\small aux}}$ typically will not have the same topology as $F$. It is also clear
that the marginal probabilities with respect to $F^{\hbox{\small aux}}$ can nevertheless (in principle)
be computed by methods analogous to those developed in Section~\ref{sec:JSF}
and Appendices~\ref{appendix1}-\ref{appendix3}, or by generalized Belief Propagation~\cite{YedidiaFreemanWeiss2003,ChertkovChernyak2006,XiaoZhou2011},
necessarily however at the cost of increased computational complexity.

\section{A two-parameter theory of 1D Ising chain dynamics}
\label{sec:2parameters}

The one-dimensional Ising chain is a convenient model since it can be solved explicitly for 
magnetizations and the pair-wise correlation functions~\cite{Glauber63}. For recent developments
on this model, for which more exact results than equal-time pair-wise correlation functions are
available, see \cite{mayer2004general}. The flip rates are
$w_i( s)=\frac{1}{2}[1-s_i\tanh[\beta h_i(s)]]$
where $h_i(s)=J (s_{i-1} + s_{i+1})$ is the local field. 
We will be interested in the relaxation
from an initial state towards equilibrium at inverse temperature $\beta$.
Periodic boundary conditions will be assumed throughout \textit{i.e.} the chain is closed.

The explicit solution for the magnetization is obtained from the exact equation
$\frac{dm}{dt}=-m+<\tanh\beta Jh>_P$ and noting that for any homogenous probability
distribution on two spins ($s_{i-1}$ and $s_{i+1}$) it reduces to
$\frac{dm}{dt}=-m(1-\tanh 2\beta J)$. It follows that for the relaxing Ising chain
any large deviation approximation of the type considered here will be exact for the magnetization because
equation (\ref{eq:mu-beta-dynamics}) is always, for the magnetization, of the type
of $-m+<\tanh\beta Jh>$.  
We note that in contrast a dynamical mean field theory gives on the ``naive mean field'' level
$\frac{dm}{dt}=-m+\tanh\left(2\beta Jm\right)$ and on the ``TAP level''
$\frac{dm}{dt}=-m+\tanh\left[(2\beta J)(m-(m+\frac{dm}{dt})J(1-m^2))\right]$, neither of which
is exact~\cite{RoudiHertz11JSAT}. 

Proceeding now to theories in $O_2$, an auxiliary Gibbs distribution based on magnetization and energy takes the form 
\begin{equation}
P_{\mbox{\ \small{Ising-2}}}({s}) = \exp\left(-\beta_M M(s)-\beta_E E(s) - F\right)
\label{eq:auxiliary-Gibbs-Ising-2}
\end{equation}
where $\beta_M$ and $\beta_E$ are the generalized temperatures at this level
of approximation and total magnetization $M$ is $\sum_i s_i$ and total energy $E$ is $\sum_i s_i s_{i+1}$.
The final state will eventually be a Boltzmann distribution at inverse temperature $\beta$,
$P\propto \exp\left(-\beta E\right)$, and the problem is hence to find out (on this level of approximation) how
($\beta_M,\beta_E$) approach ($0,\beta$), or, equivalently, how ($<m>,<e>$) approach ($0,e_{eq}$)
where $e_{eq}$ is the equilibrium energy density at temperature $1/\beta$.
The time derivatives in (\ref{eq:mu-beta-dynamics}) are   
\begin{equation}
 \frac{dm}{dt}=-m + \<\tanh\beta h\> \qquad   \frac{de}{dt}=-2e - \<h\tanh\beta h\>  
\label{eq:mu-beta-dynamics-Ising-2}   
\end{equation}
where $\<\ldots\>$ mean averages both over the chain and with respect to (\ref{eq:auxiliary-Gibbs-Ising-2}).
To compute both averages it suffices to know the marginal distributions 
$P^{\mbox{\small Ising-2}}_{j-1,j,j+1}(s_{j-1},s_j,s_{j+1})$, and in the limit of very large chain 
these can be computed by the cavity method to 
be $\propto e^{-\beta_M(s_{j-1}+s_j+s_{j+1})-\beta_E(s_{j-1}s_j+s_js_{j+1})-\tilde{h}(s_{i-1} + s_{i+1})}$
where $\tilde{h}$ is the cavity field which satisfies the fixed point equation 
\begin{equation}
\label{eq:cavfield}
\tilde{h} = \beta_M + \rm{arctanh}\left(\tanh(\beta_E \,J)\tanh(\tilde{h})\right) 
\end{equation}
The equations for the expected magnetization and the expected energy, obtained by computing the marginal
probabilities with the cavity method and then averaging are 
\begin{eqnarray} \label{eq:m-explicit}
m &=&\frac{\sinh(\beta_M + 2 \beta_E J) \mbox{e}^{2 \tilde h} + \sinh(\beta_M - 2 \beta_E J) \mbox{e}^{- 2 \tilde h} + 2\sinh{\beta_M}}{\cosh(\beta_M + 2 \beta_E J) \mbox{e}^{2 \tilde h} + \cosh(\beta_M - 2 \beta_E J) \mbox{e}^{- 2 \tilde h} + 2\cosh{\beta_M}} \\ \label{eq:e-explicit}
e &=&  \frac{-J(\sinh(\beta_M + 2 \beta_E J) \mbox{e}^{2 \tilde h} - \sinh(\beta_M - 2 \beta_E J) \mbox{e}^{- 2 \tilde h})}{\cosh(\beta_M + 2 \beta_E J) \mbox{e}^{2 \tilde h} + \cosh(\beta_M - 2 \beta_E J) \mbox{e}^{- 2 \tilde h} + 2\cosh{\beta_M}}. \\ \nonumber 
\end{eqnarray}
For this example we can also explicitly compute the marginals 
which appear in (\ref{eq:mu-beta-dynamics-Ising-2}) from (\ref{eq:auxiliary-Gibbs-Ising-2}) with the result
\begin{eqnarray}\label{eq:dm-explicit} 
&&\frac{dm}{dt} = -m + \frac{\tanh(2\beta J)[\cosh(\beta_M + 2 \beta_E J) \mbox{e}^{2 \tilde h} - \cosh(\beta_M - 2 \beta_E J) \mbox{e}^{-2 \tilde h}]}{\cosh(\beta_M + 2 \beta_E J) \mbox{e}^{2 \tilde h} + \cosh(\beta_M - 2 \beta_E J) \mbox{e}^{- 2 \tilde h} + 2\cosh{\beta_M}}; \\  \label{eq:dE-explicit} 
&&\frac{de}{dt} = -2e -\frac{2 J \tanh(2\beta J)[\cosh(\beta_M + 2 \beta_E J) \mbox{e}^{2 \tilde h} + \cosh(\beta_M - 2 \beta_E J) \mbox{e}^{-2 \tilde h}]}{\cosh(\beta_M + 2 \beta_E J) \mbox{e}^{2 \tilde h} + \cosh(\beta_M - 2 \beta_E J) \mbox{e}^{- 2 \tilde h} + 2\cosh{\beta_M}};  
\end{eqnarray}
Assuming spatial homogeneity, the dimensionality of the dynamics for this model has hence been reduced from $2^N-1$ to two. 

We implemented a routine in Mathematica which solves the time-stepping of $m$ and $e$ from (\ref{eq:dm-explicit}) and (\ref{eq:dE-explicit}) 
by a forward method and uses (\ref{eq:cavfield}), (\ref{eq:m-explicit}) and (\ref{eq:e-explicit}) to solve for $\tilde{h}$, $\beta_M$ and $\beta_E$
by Newton-Raphson. We found this routine to be stable, reflecting that we are in fact solving a discrete approximation to the
(complicated) differential equation (\ref{eq:beta-dynamics}), for which a forward method is appropriate.   
 
\begin{figure}[!htb]
\begin{center}
\includegraphics[width=6cm]{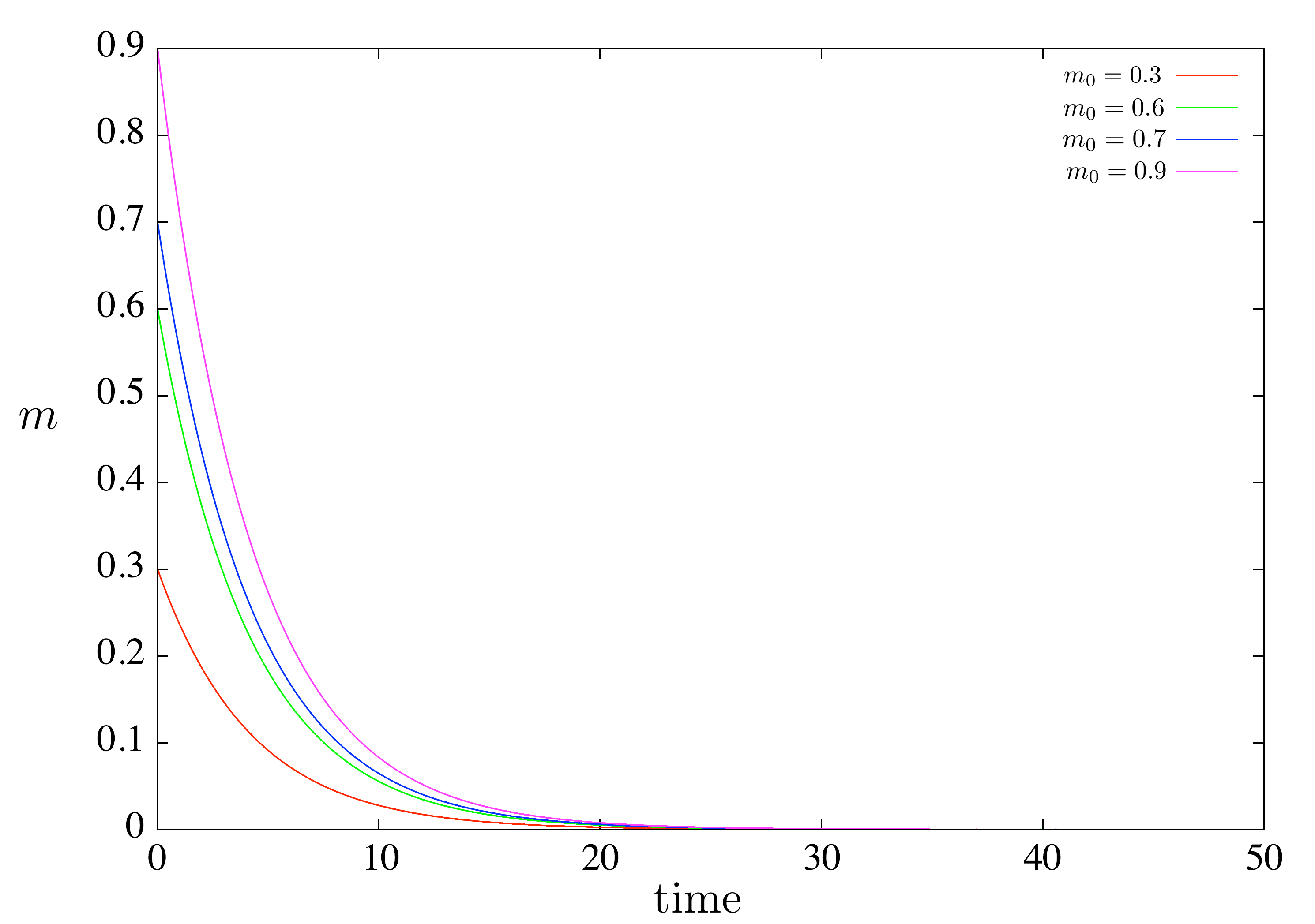}  \\
\includegraphics[width=6cm]{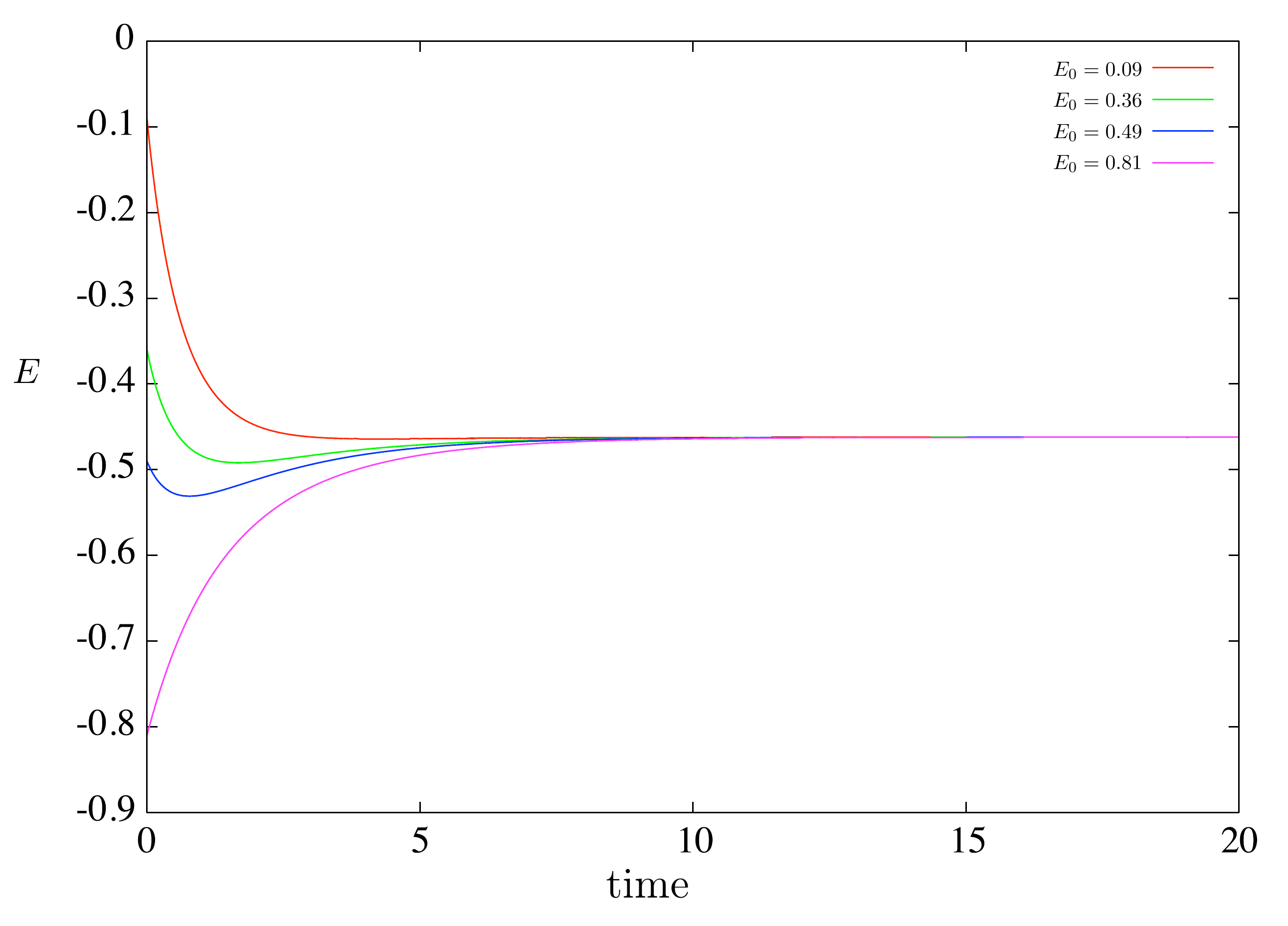}
\caption{\footnotesize Upper panel: Magnetization vs time for temperature $T=2$ and
different values of the initial conditions (initial magnetization of the system). Lower Panel: Energy vs time for same temperature and same initial conditions. 
Forward equations (\ref{eq:dm-explicit}) are integrated by using a Runge-Kutta forth order method with time step $dt= 0.001$.
Both the magnetization that the energy are observed to reach their equilibrium values, \textit{i.e.} $m_{eq}=0$ and $e_{eq}= - J\tanh(\beta \, J)$.}
\label{fig:En-2par}
\end{center}
\end{figure}

We can now compare the results of $m(t)$ and $e(t)$ to the exact results obtained by Glauber in \cite{Glauber63} which
avoids the use of Monte Carlo. For the magnetization we then have as shown above
\begin{equation}
\label{eq:mGlaub}
m(t)=m(0)\mbox{e}^{-(1-\tanh(2\beta J))\, t} \qquad 
\end{equation}
and the energy we can compute from the solutions to
Glauber's differential equation for the equal-time spin-spin correlations $r_{j,k}(t)= \ll s_i s_k \rr (t)$:
\begin{eqnarray}\label{eq:en_Glauber}
\label{eq:rGlaub}
\frac{d}{dt}\,r_{j,k}(t) = - 2 r_{j,k}(t) +\frac{1}{2}\tanh(2\beta J)\{ r_{j,k-1}(t)+r_{j,k+1}(t)+r_{j-i,k}(t)+r_{j+1,k}(t) \}.
\end{eqnarray}
Figure \ref{fig:En-2par} shows representative results of relaxation at a fixed intermediate temperature from different initial conditions
where the spins are initially independent and identically distributed as determined by the initial magnetization.
%
\begin{figure}[!ht]
\begin{center}
\includegraphics[width=6cm]{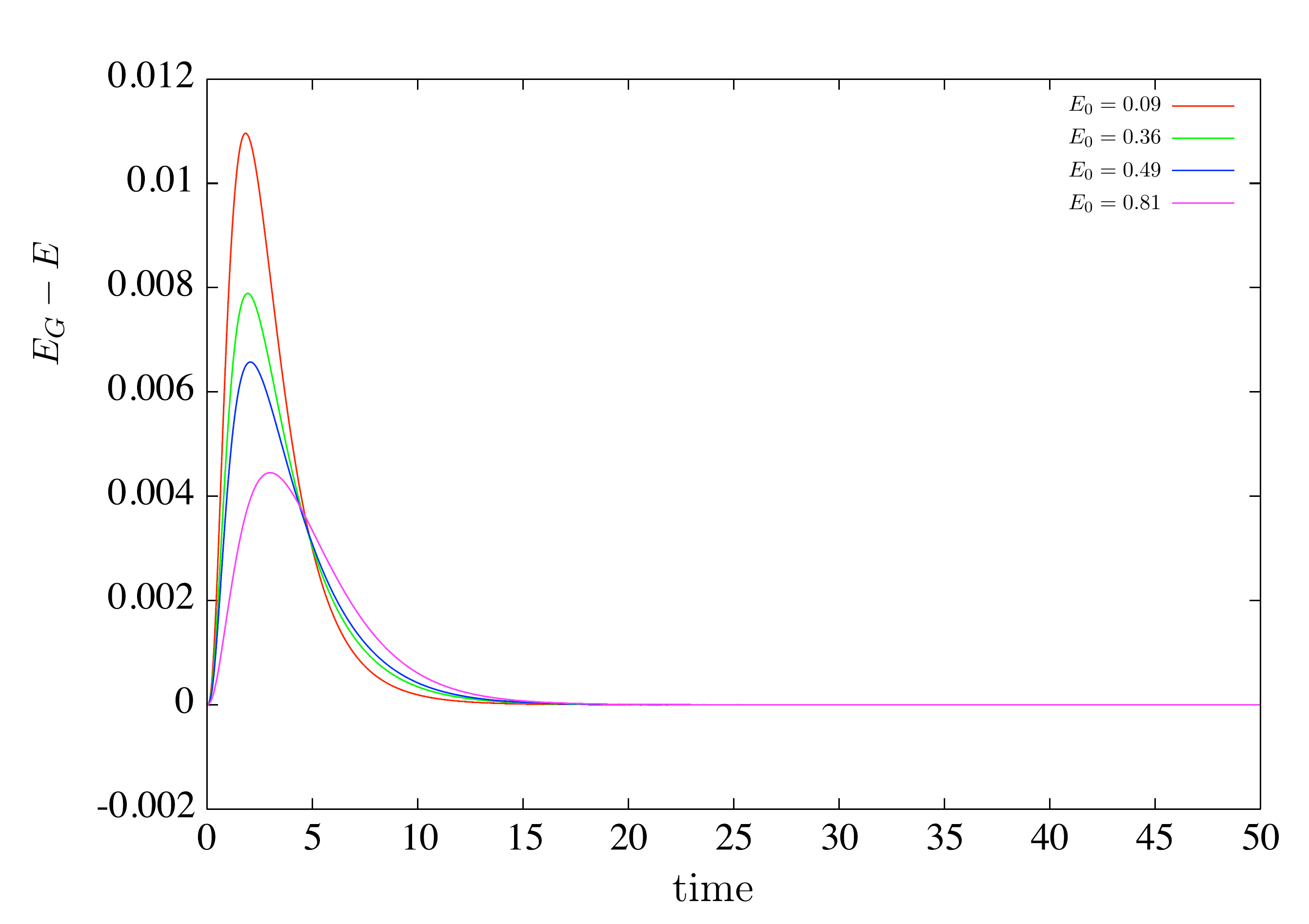}
\caption{\footnotesize Difference between the exact value of the energy computed by
Glauber theory (integrating eq (\ref{eq:en_Glauber}) with a number of spins $N=10^5$) and the approximate two-parameter theory developed in this section. 
Temperature $T=2$, different curves are referred to different value of the initial condition. It is observed that the deviations are small initially (as
follows from the chosen initial conditions), as well as at long terms (when equilibrium is reached). Discrepancies are instead found
at intermediate times.}
\label{fig:En_diff_2par}
\end{center}
\end{figure}
The results obtained for the magnetization shown in the upper panel of Fig.~\ref{fig:En-2par} 
are of course in perfect agreement with the prediction of the Glauber theory, and will not be discussed
further. The lower panel of Fig.~\ref{fig:En-2par} shows the energy starting from the same initial conditions. 
These curves are in qualitative agreement with the Glauber theory
at early and late times, but shows a discrepancy at intermediate times.
In this region, where the time derivative of the energy changes sign, the spins are correlated over a larger distance
effects which are not included here by the probability distribution (\ref{eq:auxiliary-Gibbs-Ising-2}).

\begin{figure}[!ht]
\begin{center}
\includegraphics[width=6cm]{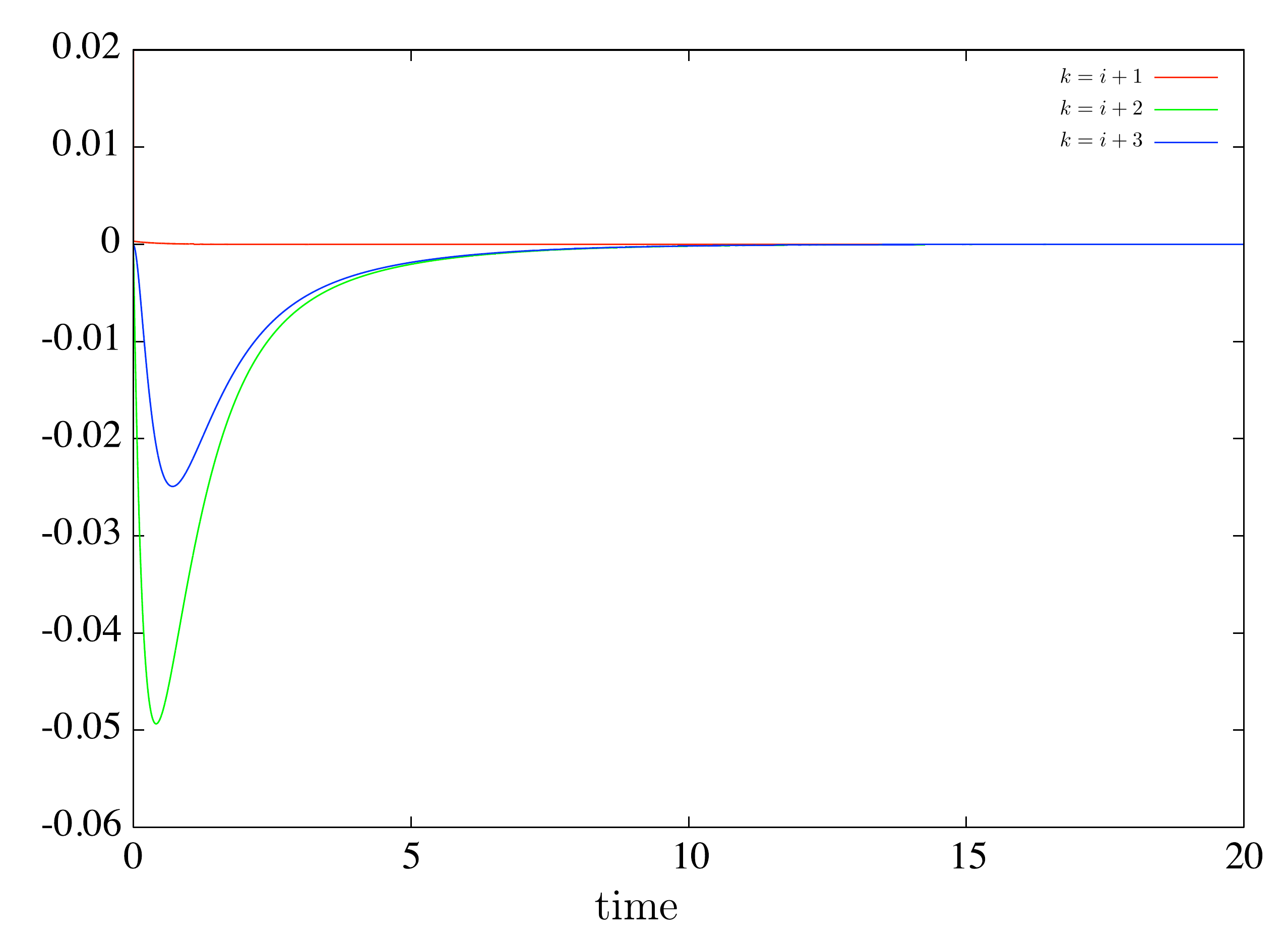}  
\caption{\footnotesize The difference $\frac{d^{(T)}}{dt} \ll  s_i s_k \rr -\frac{d^{(M)}}{dt} \ll  s_i s_k \rr $, taken from 
(\ref{eq:deriv-corr-true}) 
and (\ref{eq:deriv-corr-projected}), is shown for the index $k$ taking the values $k=\{i+1, i+2, i+3\}$. The time derivative of the correlation function $\ll s_i s_{i+1} \rr$ is perfectly recovered with the presented approach (red line), while the time derivative of the correlation  $\ll s_i s_{i+2} \rr$ and $\ll s_i s_{i+3} \rr$ (green and blue line respectively), correlation functions which are not included in the model, are not well reconstructed at intermediate times. The latter difference (blue line) appears to be smaller simply because correlations between spins at longer distance are weaker.} 
\label{fig:Corr-diff-2par}
\end{center}
\end{figure}

We now turn to an internal test of the theory by considering in (\ref{eq:error-Q})   
pair-wise correlation functions of spins which are not necessarily neighbors,
\emph{i.e.} $\ll s_i s_k \rr$. Recall that we want to estimate these quantities
not for the exact dynamics, for which they are (for the Ising chain)
already given by the Glauber theory (\ref{eq:en_Glauber}),
but how they differ when the probability distribution is
at all time taken to be (\ref{eq:auxiliary-Gibbs-Ising-2}).
We then have for one term
\begin{align}\label{eq:deriv-corr-true}
\frac{d^{(T)}\ll s_i s_k \rr }{dt t} &= \frac{\partial}{\partial t} \sum_{ s} P(s)\, s_i s_k = \sum_{ s} \frac{\partial}{\partial t} P(s)\, s_i s_k \\ \nonumber
 &= \sum_{s} \sum_{j=1}^N \{w_j(F_j s)P(F_j s)-  w_j(s) P(s) \} s_i s_k = -2 \sum_{s} P(s)\, \{w_i({s})+ w_k({s})\}
\end{align}
and for the other 
\begin{align}\label{eq:deriv-corr-projected}
\frac{d^{(M)}\ll s_i s_k \rr }{d t} &= \dot{\beta}_M \big( \ll M(s_i s_k)\rr - \ll M\rr \ll s_i s_k \rr\big)+\dot{\beta}_E \big( \ll E(s_i s_k)\rr - \ll E\rr \ll s_i s_k \rr\big)
\end{align}
As shown in Figure~\ref{fig:Corr-diff-2par} the time change of the correlation function $\ll  s_i s_{i+1} \rr$ (the energy), which is included in the
two-parameter theory, is reproduced exactly, while those of  $\ll  s_i s_{i+2} \rr$ and $\ll  s_i s_{i+3} \rr$ are not.
This then shows, internal to the theory and without solving the full dynamical equations, that the model does not
catch effective interactions which develop between non-neighboring spins at intermediate times, which is
also where we find the largest discrepancies between the energy computed exactly (by Glauber's equation)
and in the model, compare Fig.~(\ref{fig:En_diff_2par}).

%

\section{A joint spin-field theory of 1D Ising chain dynamics}
\label{sec:JSF}
We start this section by noting generally that if the master equation describes a process obeying local balance
then the flip rate $w_j$ is determined (and only determined) by the value of $s_j$ and the total local field acting on spin $j$,
$h_j(s)=\sum_l (O_l(F_js)-O_l(s))/(2s_j)$. For the example of a relaxing Ising spin chain $h_j(s)=J\left(s_{j-1}+s_{j+1}\right)$.
At least for systems obeying detailed balance it is therefore reasonable to assume that a description in terms of spins 
and total local fields acting on these spins could be accurate.
Laughton et al in~\cite{DRTinf2} were the first to investigate this possibility, and in the terminology used
here can be said to have proposed to use  
\begin{equation}
P^{\hbox{\small LCS}}= \exp\left(\sum_{s,h,i}d(s,h)1_{s_i,s}1_{h_i,h}-F\right)
\label{eq:auxiliary-LCS}
\end{equation}
This is called the ``joint spin-field distribution'' model as the
measure in (\ref{eq:auxiliary-LCS})
simply counts the number $n_{s,h}$ of spins
taking values $s$ and having total local fields $h$,
and weighs the total spin configuration by $\prod_{s,h}e^{ d(s,h)n_{s,h}}$. 
If implemented naively this leads to factor graph $F^{\hbox{\small aux}}$ which is not locally tree-like since the 
fields and spins overlap. For an Ising spin chain a function of $s_j$ and $h_j$ actually depends on $(s_{j-1},s_j,s_{j+1})$,
a function of $s_{j+1}$ and $h_{j+1}$ depends on $(s_{j},s_{j+1},s_{j+2})$, leading to staggered auxiliary
energy functions along the chain. A further issue is that the parametrization of the probability
distribution in (\ref{eq:auxiliary-LCS}) is over-complete. 

\begin{figure}[!ht]
\begin{center}
\includegraphics[width=6cm]{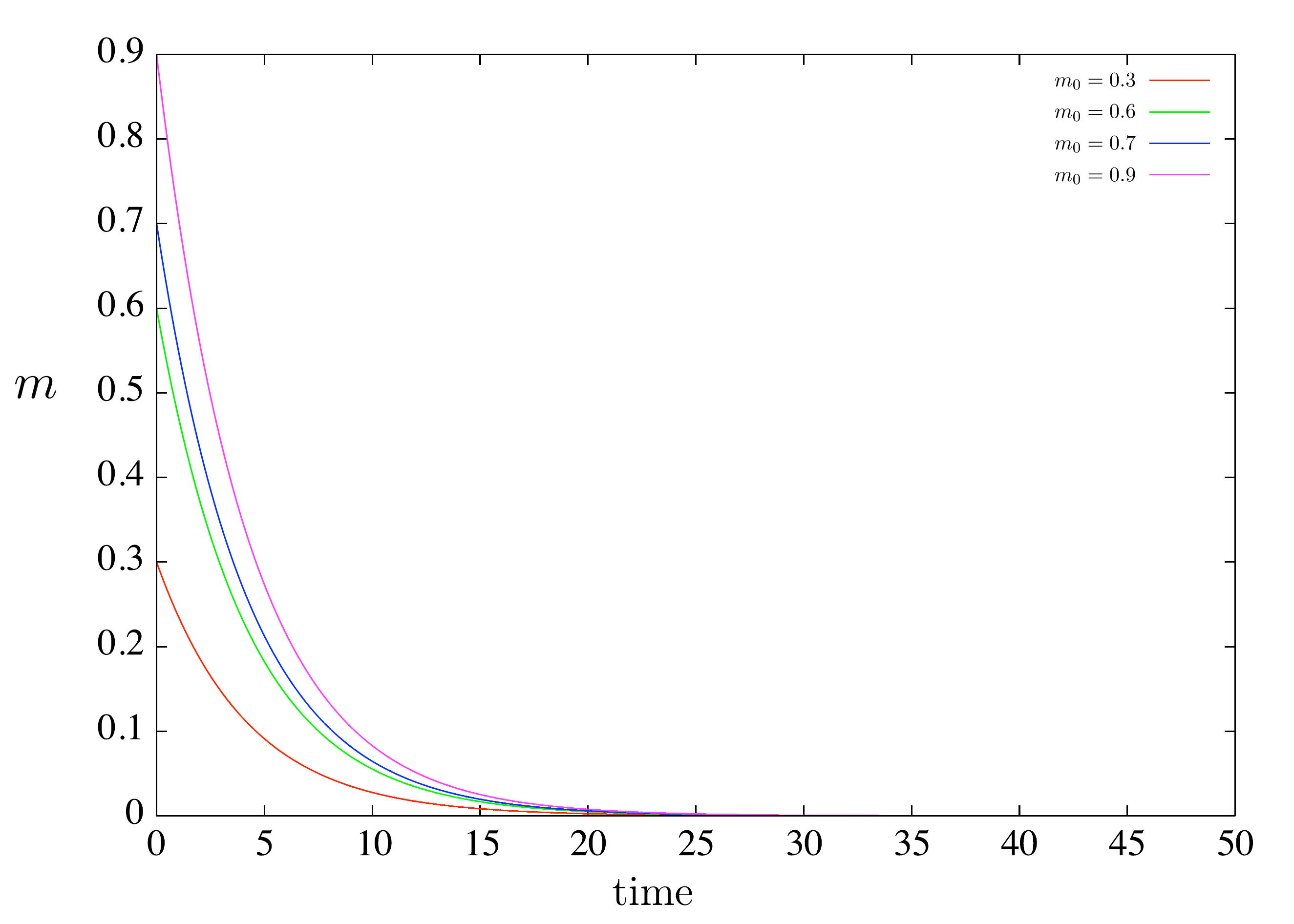}   \\
\includegraphics[width=6cm]{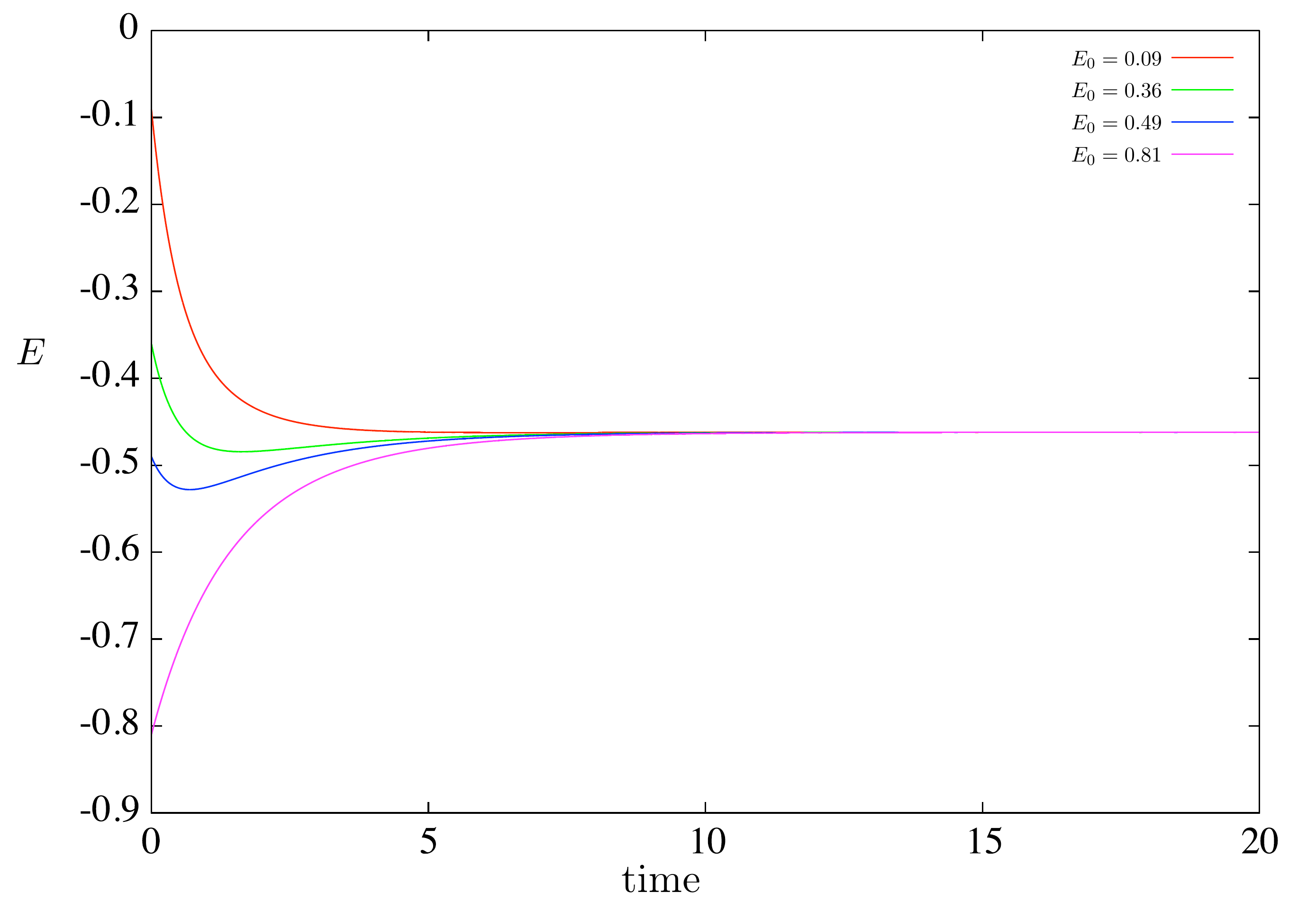}
\caption{\footnotesize Upper panel: Magnetization vs time for different value of the initial conditions (initial magnetization of the system) obtained by using the joint spin-field distribution dynamics (\ref{eq:Dsh-dynamics}). Lower Panel: Energy vs time for different initial conditions obtained with the same method. In both plots the temperature $T=2$.Magnetization and energy reach their equilibrium values \textit{i.e.} $m_{eq}=0$ and $e_{eq}= - J\tanh(\beta \, J)$. Qualitatively the curves are
similar energy vs time curve is closer to the Glauber theory than in the two-parameter theory of Section~\protect\ref{sec:2parameters}, 
see Fig.~\protect\ref{fig:En-2par}.
}
\label{fig:En_Dsh}
\end{center}
\end{figure}

\begin{figure}[!ht]
\begin{center}
\includegraphics[width=6cm]{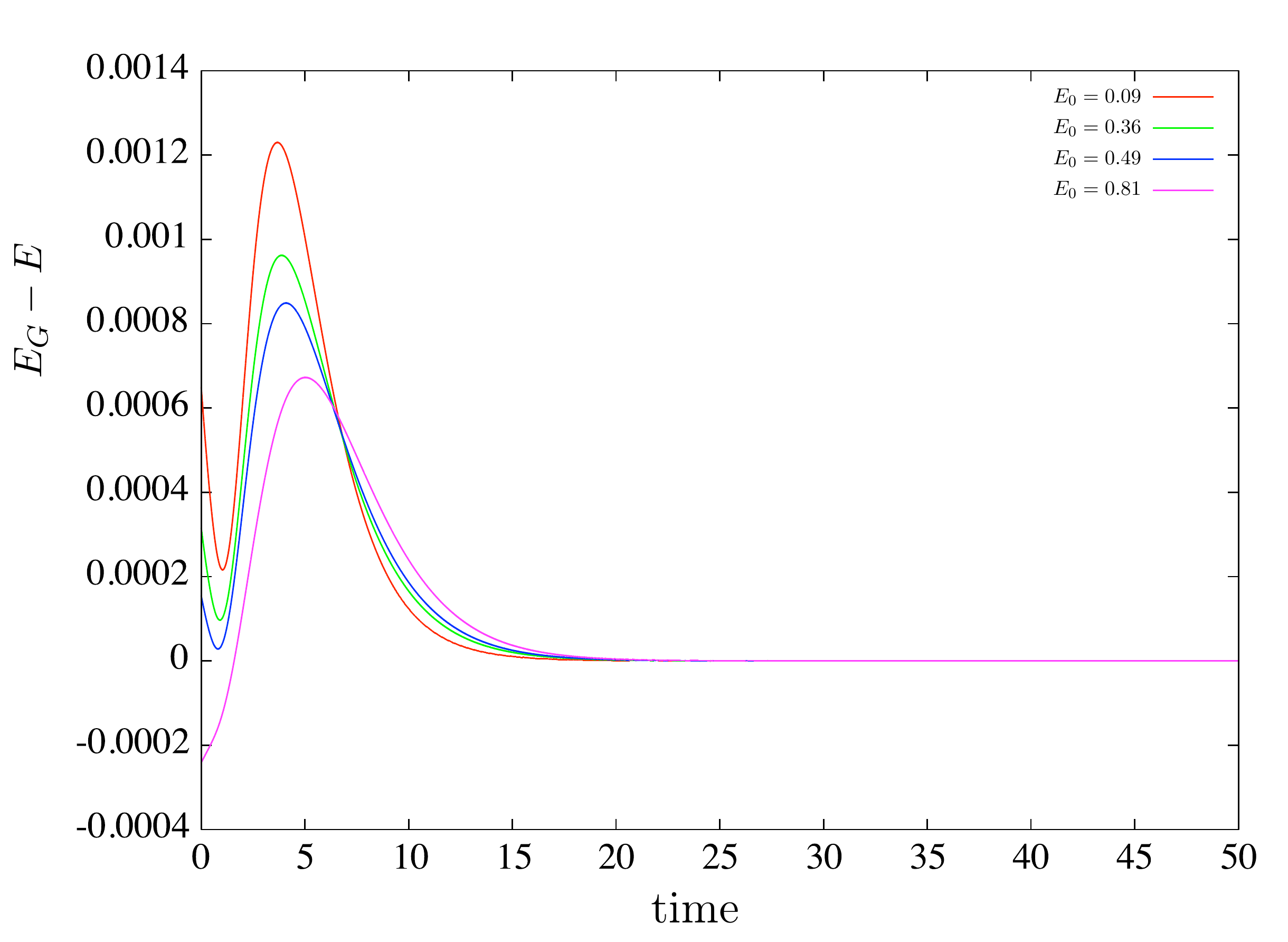}  
\caption{\footnotesize Difference between the energy computed by the Glauber theory (integrating eq (\ref{eq:en_Glauber}) with a number of spins $N=10^5$) and by the joint spin-field theory. 
Different curves refer to different values of the initial conditions, temperature is $T=2$. The results agree to one order of magnitude better than the ones obtained in Section \ref{sec:2parameters}, see Fig.~\ref{fig:En_diff_2par}. 
}
\label{fig:En-diff-Dsh}
\end{center}
\end{figure}

We will now show how the issue of  $F^{\hbox{\small aux}}$ can be addressed by
introducing ancillary dummy variables such that the joint spin-field distribution model can 
be formulated on an auxiliary locally tree-like factor graph and the necessary averages evaluated by Belief Propagation.
This in fact allows for somewhat more general theories than the joint spin-field distribution (at the same level of model complexity).
We will carry out the argument for general pairwise interactions, and only at the end specialize to the one-dimensional chain. 
We start by inflating the factor graph and changing any link between variable $i$ and factor $a$ into 
a new variable node $ia$ holding the variable $(s_{ia},h_{ia})$. This variable node is
connected to precisely two factor nodes in the inflated graph corresponding to old factor node $a$ and old variable node $i$.
The first component of $(s_{ia},h_{ia})$
is a spin variable (``spin $i$ as factor $a$ thinks it is'') and the second component is a field
(``local field on spin $i$ from factor $a$, as factor $a$ thinks it is''). This field $h_{ia}$ takes values
in a discrete set which are the values
that a local field on spin $i$ from the energy term $\epsilon_a$ can take in the original problem. 
A factor node $a$ remains a factor node in the inflated graph
and holds the constraints $h_{ia}=(\epsilon_a(F_is^a)-\epsilon_a(s^a))/2 s_{ia}$ where now 
$s^a$ stands for the collection $\{s_{ia}\}_{i\in\partial a}$. New factor node $a$ 
is also allowed to hold any other function $A_a$ of the collection $\{(s_{ia},h_{ia})\}_{i\in\partial a}$.
A variable node $i$ is on the other hand changed into a factor node in the inflated graph and holds the equality constraint
$s_{ia}=s_{ib}=\ldots$. 
New factor node $i$ is also allowed to hold any function
$B_i$ of the collection $\{(s_{ia},h_{ia})\}_{a\in\partial i}$. 
Finally we can allow any functions $C_{ia}$ of the variables $(s_{ia},h_{ia})$ themselves (``external fields acting on the dummy variables'').
The auxiliary Gibbs distribution of the new model is then
\begin{align}
P({s,h}) =& \, e^{-F}\prod_iB_{i} 1_{s_{ia}=s_{ib}=\ldots = \overline{s}_i} \prod_{ia}C_{ia}  
 \prod_aA_{a}\prod_{i\in\partial a}1_{h_{ia},(\epsilon_a(F_is^a)-\epsilon_a(s^a))/2s_{ia}}
\label{eq:auxiliary-LCS-dummy}
\end{align}
It is clear that (\ref{eq:auxiliary-LCS-dummy}) will be equivalent to (\ref{eq:auxiliary-LCS}) if
$A_{a}=C_{ia}=1$ and $B_{i}=\exp\left(\sum_{s,h}d(s,h)1_{s,\overline{s_{i}}}1_{h,\sum_a h_{ia}}\right)$
where $\overline{s_{i}}$ can be taken any one in the set $\{s_{ia}\}$ since the constraints mean that these
dummy variables all must take the same value.
On the other hand, the factor graph describing (\ref{eq:auxiliary-LCS-dummy}) inherits the
topology of the original factor graph, and if the original factor graph is locally tree-like,
so is the factor graph describing (\ref{eq:auxiliary-LCS-dummy}). 
The observables conjugate to the $d(s,h)$ are
\be\label{eq:JSF}
 \mu(s',h') = \frac{1}{N}\sum_i \<1_{s',\overline{s_{i}}}1_{h',\sum_a h_{ia}}\>.
\ee
The parametrization (\ref{eq:auxiliary-LCS}) is over-complete because $\sum_{s',h'} \mu(s',h')=1$.
If we want to compare to the two-parameter theory the
magnetization and energy can be computed from (\ref{eq:JSF}) as
\be
m_t= \sum_{s',h'} \mu_t(s',h') s',  \qquad e_t = -\frac{1}{2} \sum_{s',h'} \mu_t(s',h') s' \, h'
\ee
where the subscript $t$ indicates the time.
We show in~\ref{appendix1} that for the Ising chain model and the joint spin-field distribution
the general equation (\ref{eq:mu-beta-dynamics}) reads
\begin{align} \nonumber
 \frac{d\mu(s',h')}{dt}=&\, \frac{1}{2}\left(1+s' \tanh(\beta h')\right)\mu(-s',h') - \frac{1}{2}\left(1-s'\tanh(\beta h')\right)\mu(s',h') \\ \nonumber
 +&\frac{1}{\Omega} \sum_{s_{j},s_{j+1}} \sum_{s_{j-1}, s_{j+2}}\,\, \mbox{e}^{d(s_{j},J (s_{j+1} + s_{j-1}))} \mbox{e}^{d(s_{j+1},J(s_{j} + s_{j+2} ))} \frac 1 2 (1 - s_{j}\tanh(\beta h_j{(s)}))  \\
&  \mbox{e}^{\theta J (s_{j}+s_{j+1})+\nu (s_{j-1}+s_{j+2}) +\eta J (s_{j}s_{j-1} + s_{j+1}s_{j+2})} [1_{s',\bar s_i}(1_{h', -Js_j +J s_{j+2}}-1_{h',J(s_j + s_{j+2})})]
\label{eq:Dsh-dynamics}
\end{align}
where $\theta, \nu$ and $\eta$ are cavity fields which are needed to compute the marginal probabilities
in the auxiliary factor graph, and which satisfy the system of equations (\ref{eq:cavity-field}). 
In Appendix~\ref{appendix2} we show that the same results can obtained using a version of the cavity method
which is not explicitly reduced to ordinary BP, and in Appendix~\ref{appendix3} we show that the two approaches both lead to (\ref{eq:Dsh-dynamics}).
\\
To numerically obtain the dynamics of the joint spin-field distribution one should solve the time-stepping of $\mu(s',h')$ from (\ref{eq:Dsh-dynamics}) and then use (\ref{eq:JSF}) and the equations (\ref{eq:cavity-field}) to get the values at time $t$ of $d(s',h')$ and of the cavity fields $\nu, \theta, \eta$  by Netwon-Raphson, where the average in (\ref{eq:JSF}) can be taken by using the cavity method, \emph{i.e.} $P(s,h) \propto \exp{(d(s,h) + \nu s_i + \theta J s_{i+1}+\eta J s_i s_{i+1})}$. Unfortunately, because of the constraint $\sum_{s',h'} \mu(s',h')=1$ the equations (\ref{eq:JSF}) are not independent and then the system of 9 equations made by (\ref{eq:JSF}) and (\ref{eq:cavity-field}) which has to be solved has singular Jacobian. We solved this problem by making an observation that allows us to invert the equations (\ref{eq:JSF}) respect to the parameters $d(s',h')$. Let us note that in general these equations are not invertible because, for every couple $(s',h')$, the joint spin-field distribution depends on the 3 cavity fields and on the all 6 parameters $d(s',h')$. Nevertheless observing that this full dependence comes mainly from the partition function which contains all the $d(s,h)$ terms, we can consider a not normalized version of (\ref{eq:JSF}) (where every $\mu(s',h')$ still depends on the 3 cavity fields but on just one $d(s',h')$: that one with the same configuration of $(s',h')$) and then invert them to get $d(s',h')$ as a function of $\{\mu_{n-n}(s',h'), \nu, \theta, \eta\}$, where the label $n$-$n$ here means ``not-normalized''. Let us note that, made this inversion, also the cavity fields are expressed as functions of the cavity field them selves and of the $n$-$n$ joint-spin field distribution instead of the $d(s,h)$ parameters. Once this has been performed, given initial values for the $\mu_{n-n}(s',h')$'s which are analytically computable, it is possible to get the values of $\nu, \theta, \eta$ by Newton-Raphson by using these new expressions of the cavity field equations. Then the $d(s',h')$ can be computed from their analytical expression and the $\mu(s',h')$'s can be normalized afterwards to solve the time-stepping by  (\ref{eq:Dsh-dynamics}) and iterate the procedure. We conclude observing that this scheme also reduces the dimensionality of the system of equations which has to be solved by Newton-Raphson from 9 to 3 equations.
\\
The results for the magnetization and energy obtained with this procedure are shown in figure \ref {fig:En_Dsh}, \ref{fig:En-diff-Dsh} and can be compared with those in Section \ref{sec:2parameters}. As one can see, using the joint spin-field distribution approach improves by one order of magnitude the agreement
to the Glauber theory. As can be expected from a theory which includes general terms in $O_2$ we capture (locally) correctly both
the time change of the nearest-neighbor correlation function
$\ll s_i s_{i+1}\rr(t)$ (as already did the two-parameter theory) but also the next-nearest neighbor correlation functions $\ll s_i s_{i+2}\rr(t)$. 
More distant correlations are however still not exactly reconstructed by the $\mu(s',h')$ theory, in agreement with the general perturbative scheme 
worked out in Section~\ref{sec:perturbation}, see caption to Fig.~\ref{fig:En-diff-Dsh}
%
\begin{figure}[!ht]
\begin{center}
\includegraphics[width=6cm]{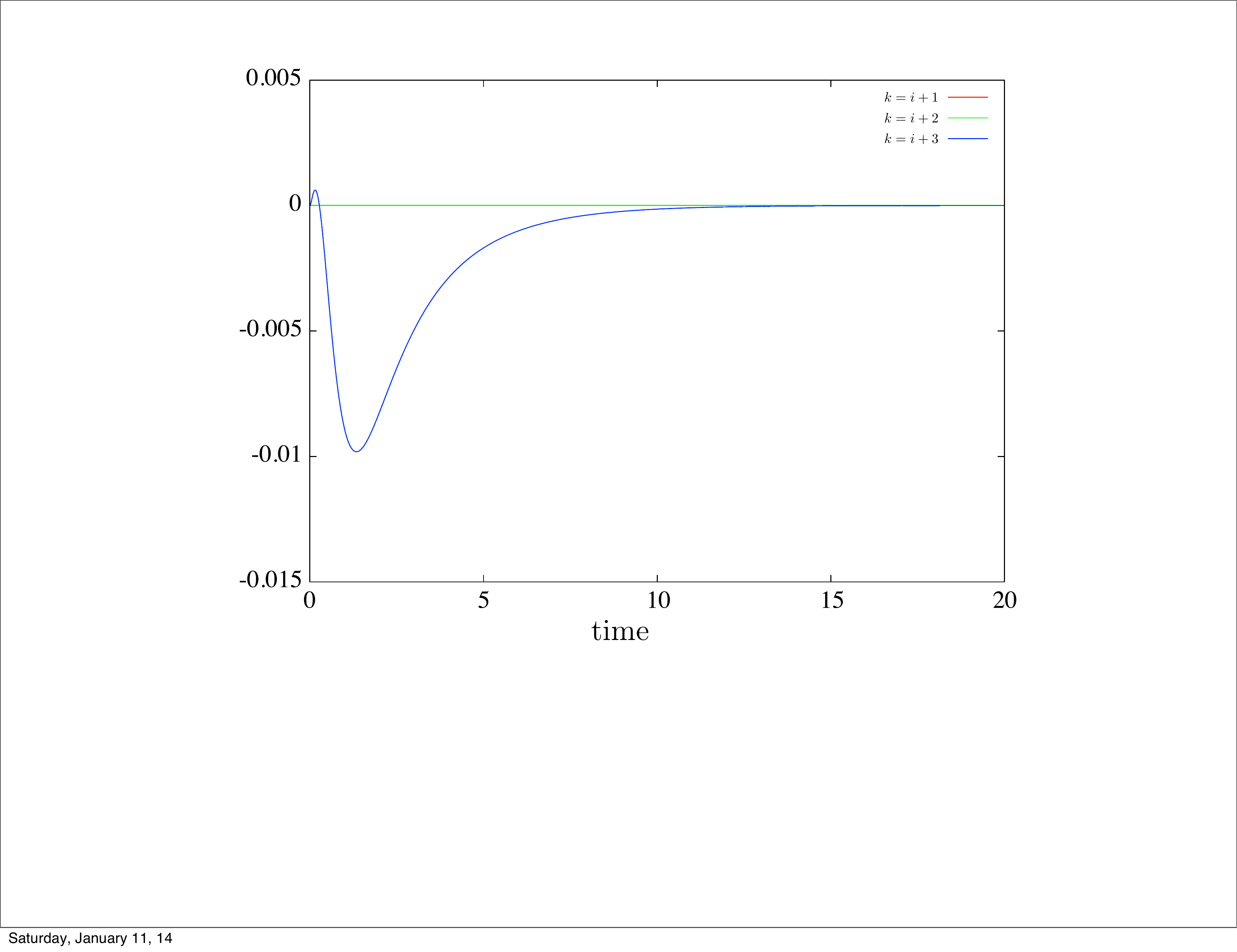}
\caption{\footnotesize The differences $\frac{d^{(T)}}{dt} \ll  s_i s_k \rr -\frac{d^{(M)}}{dt} \ll  s_i s_k \rr $ for $k=\{i+1, i+2, i+3\}$. Compared to Fig.~(\protect\ref{fig:Corr-diff-2par}), the joint spin-field theory reproduces the time derivatives of both correlation function $\ll s_i s_{i+1} \rr$ and $\ll s_i s_{i+2} \rr$  (red and green line overlapping in the figure), while the time derivative of correlation $\ll s_i s_{i+3} \rr$ (blue line), being outside our parametrization, is not well reconstructed in the most off-equilibrium region.}
\label{fig:Dsh_correlations}
\end{center}
\end{figure}
\noindent

\section{Discussion}
\label{sec:discussion}
In this paper we have pointed out 
that if the probability distribution of a non-equilibrium system obeys a large deviation
then this can be combined with methods to efficiently compute marginals
of Gibbs distributions developed in disordered systems theory~\cite{MMbook}
together entailing
a very considerable dimensional reduction of a spin system dynamics described by a master equation.
We have also pointed out that the accuracy
of such a dimensional reduction can be assessed self-consistently without reverting to
a simulation of the full system.
These tests of self-consistency amount to computing the time change of correlation functions
which are not in the assumed large deviation principle in two different ways,
and which have to agree if the large deviation principle is an accurate approximation. As far as we are aware,
such tests have previously only been carried out by Nishimori and Yamana in~\cite{NishimoriYamana1996},
in the specific setting of a high-temperature expansion of the dynamic SK model, see curve of ``$c_3(t)$'' in Fig.~1 of~\cite{NishimoriYamana1996}.
We believe that such tests are in fact central to the validity and usefulness of the approach. 
In Appendix~\ref{appendix4} we sketch a geometrical interpretation of the reduction as a projection on 
hierarchies of probability distribution, a concept developed in information theory~\cite{Amari2001}.

The general scheme presented here may be conceptually important as
a possible basis for a perturbation scheme for non-equilibrium systems akin to cluster expansions.
One of the more promising potential applications could be to describe the puzzling behavior
of focused local search on large random satisfiability problems (physically an instance
of a bulk-driven non-equilibrium process)~\cite{Alava2008}, which is of considerable practical importance,
and which has so far defied theoretical understanding beyond a special case investigated
already a decade ago~\cite{BarthelHartmannWeigt2003,MonassonSemerjian2003}.

A serious limitation of the approach, as applied to real physical systems, which likely has not escaped the reader, 
is that we have assumed
spatial homogeneity. If a non-equilibrium system varies in space then 
a large deviation principle described by an auxiliary Gibbs distribution would need (auxiliary)
energy terms which also vary in space, necessarily adding to the model complexity.
Furthermore, it was shown long ago~\cite{KirkpatrickCohenDorfman1982}, and is well established 
experimentally~\cite{LawSengers1989}, that non-equilibrium states driven by spatially varying boundary conditions and
exhibiting fluxes generally display long-range correlation functions which
in macroscopic fluctuation theory have been shown to correspond to long-range effective
interactions (``non-local entropy functionals''), see~\cite{Derrida2007}.
For systems extended in space the approach would therefore seem to be limited to relaxation towards
a steady steady state ($\frac{\delta S}{\delta \rho}=0$ in the language of~\cite{Derrida2007}), 
while the more interesting case of spontaneous fluctuations away from a steady state
($\frac{\delta S}{\delta \rho}\neq 0$) would be out of reach.

\section{Acknowledgements}
The authors warmly thank Dr. A. Mozeika for suggesting this project and for very useful discussions. The approach developed below in Appendix B was shown to one of the author (GDF) by Dr. Mozeika.
This work has been supported by the People Programme (Marie Curie Actions) of the European Union's Seventh Framework Programme FP7/2007-2013/ under REA grant agreement n. 290038 (GDF), by the Academy of Finland as part of its Finland Distinguished Professor
program, project 129024/Aurell, and through the Academy of Finland Center of Excellence COIN (EA). 
EA thanks the organizers of ICSG2013 (Sapporo, Japan, July 2013) for an opportunity to
present part of this work, and Yoshiyuki Kabashima for valuable discussions.


\appendix 

\section{Joint spin-field distribution formalism by using a large deviation approach} 
\label{appendix1}
The starting point is the auxiliary probability distribution in (\ref{eq:auxiliary-LCS-dummy}).
The factor graph corresponding to this model has the same topology as the original graph
but is inflated, \textit{i.e.} new variables nodes have been introduced corresponding to
the links in the original graph.
In the general Belief Propagation formalism one would here need to consider four kinds of message
$m_{i\to ia}$, $m_{a\to ia}$, $n_{ia\to i}$, and $n_{ia\to a}$, each of them a probability distribution over $(s_{ia},h_{ia})$~\cite{MMbook}.
However, since each new variable node $ia$ is connected to only two new factor nodes and since 
$C_{ia}=1$ we have $n_{ia\to i}=m_{a\to ia}$ and $n_{ia\to a}=m_{i\to ia}$, reducing the number of types of messages to two.
Furthermore, for pairwise Ising interactions the possible values of $h_{ia}$ can be parametrized as $J\tau_{ia}$ where
$\tau_{ia}$ is another spin variable, and one of the two Belief Propagation update equations reads
\begin{eqnarray}
 m_{a\to ia}(s_{ia},\tau_{ia})&\propto&\sum_{(s^a,\tau^a)\setminus (s_{ia},\tau_{ia})}\prod_{j \in \partial a} 1_{J\tau_{ja},\epsilon_a(F_js^a)-\epsilon_a(s^a)/2s_{ja}} \nonumber \\
                        &&\quad \prod_{j\in\partial a \setminus i} m_{j\to ja}(s_{ja},\tau_{ja})
\label{eq:BP-dummy-1}
\end{eqnarray}
Since the factor node $a$ can be identified as the pair $(i,j)$ the sum in above is in fact only over $(s_{ja},\tau_{ja})$
and the first product enforces the two constraints $J\tau_{ia}=Js_{ja}$ and $J\tau_{ja}=Js_{ia}$
which give $s_{ja}=\tau_{ia}$ and $\tau_{ja}=s_{ia}$.
Introducing the simpler notation $s_{ja}=s_{ji}$, $\tau_{ja}=\tau_{ji}$ and $m_{j\to ja}(s_{ja},\tau_{ja})=\mu_{j\to i}(s_{ji},\tau_{ji})$
we can therefore write (\ref{eq:BP-dummy-1}) as the identity
\begin{equation}
 m_{a\to ia}(s_{ij},\tau_{ij})=\mu_{j\to i}(\tau_{ij},s_{ij}).
\label{eq:BP-dummy-2}
\end{equation}
This reduces the number of types of messages to one. The other side of the Belief Propagation update equations then reads
\begin{eqnarray}
 \mu_{i\to j}(s_{ij},\tau_{ij})&\propto& \sum_{\tau^i\setminus \tau_{ij}}e^{d\left(s_{ij},\, J\sum_{k\in\partial i}\tau_{ik}\right)}\nonumber \\
&&\quad \prod_{k\in\partial i\setminus j}\mu_{k\to i}(\tau_{ik},s_{ij}) 
\label{eq:BP-dummy-3}
\end{eqnarray}
where the sum over the variables $s^i\setminus s_{ij}$ has been carried out using equality constraint
in (\ref{eq:auxiliary-LCS-dummy}).
For the Ising spin chain where spin $i$ only interacts with spins  $i- \equiv i-1$ and $i+\equiv i+1$ the sum in (\ref{eq:BP-dummy-3}) is 
only over one term and (\ref{eq:BP-dummy-3}) is further reduced to
\begin{eqnarray}\nonumber
 \mu_{i\to (i+1)}(s_{i+},\tau_{i+})&\propto& \sum_{\tau_{i-}}e^{d\left(s_{i+},\,J(\tau_{i-}+\tau_{i+})\right)} \\ 
 &&\mu_{(i-1)\to i}(\tau_{i-},s_{i+})
\label{eq:BP-dummy-4}
\end{eqnarray}
and analogously in the other direction. 
Since a probability distribution on two spin variables can be written
$\mu_{i\to (i+1)}(s_{i},\tau_{i})\propto \exp\left(\theta_{i} J s_{i}+\nu_{i}\tau_{i}+\eta_{i}J s_{i}\tau_{i}\right)$ we can
rewrite the previous equation as:
\begin{align}\label{eq:mess-exp} \nonumber
e^{\theta J s_{i+}+\nu \tau_{i+} +\eta J s_{i+}\tau_{i+}} &\propto \sum_{\tau_{i-}} e^{d\left(s_{i+},\, J(\tau_{i-}+\tau_{i+})\right)}\\
& e^{\theta J \tau_{i-}+\nu s_{i+}+\eta J \tau_{i-}s_{i+}}
\end{align}
(where we implicitly assumed homogeneity, \emph{i.e.} $\theta_i=\theta, \nu_i=\nu$ and $\eta_i=\eta$) which, for a large chain, can be taken as a fixed point equation for three cavity fields $\theta$, $\nu$ and $\eta$
generalizing the simpler expression in (\ref{eq:cavfield}) on the original factor graph.
These fixed point equations read as follow:
 \begin{align}\label{eq:cavity-field}\nonumber
\nu &= \frac 1 4 \log{\Bigg[ \frac{(\rm{e}^{-\eta + \theta- \nu + d(1, -2)} + \rm{e}^{\,\eta + \theta + \nu + d(1, 0)})(\rm{e}^{\,-\eta + \theta - \nu + d(1, 0)} + \rm{e}^{\,\eta + \theta + \nu + d(1, 2)})}{(\rm{e}^{\,\eta - \theta - \nu+ d(-1, -2)} + \rm{e}^{\,-\eta-\theta+\nu + d(-1, 0)})(\rm{e}^{\,\eta - \theta - \nu+ d(-1, 0)} + \rm{e}^{\,-\eta - \theta + \nu + d(-1, 2)})}\Bigg]}\\ 
\theta&= \frac 1 4 \log{\Bigg[\frac{(\rm{e}^{\,\eta - \theta - \nu + d(-1, 0)} + \rm{e}^{\,-\eta - \theta + \nu + d(-1, 2)}) (\rm{e}^{\,-\eta + \theta - \nu + d(1, 0)} +  \rm{e}^{\,\eta + \theta + \nu + d(1, 2)})}{(\rm{e}^{\,\eta - \theta - \nu + d(-1, -2)} + \rm{e}^{\,-\eta - \theta + \nu + d(-1, 0)})(\rm{e}^{\,-\eta + \theta - \nu + d(1, -2)} + \rm{e}^{\,\eta + \theta + \nu + d(1, 0)})}\Bigg]} \\
 \eta&=\frac 1 4 \log{\Bigg[ \frac{(\rm{e}^{\,\eta - \theta - \nu + d(-1, -2)} + \rm{e}^{\,-\eta - \theta + \nu + d(-1, 0)})(\rm{e}^{\,-\eta + \theta - \nu + d(1, 0)} +  \rm{e}^{\,\eta + \theta + \nu + d(1, 2)})}{(\rm{e}^{\,\eta - \theta - \nu + d(-1, 0)} + \rm{e}^{\,-\eta - \theta + \nu + d(-1, 2)}) (\rm{e}^{\,-\eta + \theta - \nu + d(1, -2)} + \rm{e}^{\,\eta + \theta + \nu + d(1, 0)})} \Bigg]} \nonumber
\end{align}\\
At this point we have to reinterpret the flip operations $F_j$ in (\ref{eq:master-equation}) to mean 
an event labeled $j$ where 
all dummy spin variables $s_{ja}$ are flipped simultaneously and all dummy local fields 
$h_{ia}$ where $a\in\partial j$ are also changed according to a changed value of $s_{ja}$
(we temporarily revert to the formulation in terms of $\{(s_{ia},h_{ia})\}$).
In addition we assume that the dependence of rate $w_j$
depends as before on a total local field, but which is now defined as $\sum_a h_{ja}$.
If so, we can write
the master equation for a probability distribution over the dummy variables as    
\begin{align}
 \partial_tP(s,h) = \sum_{j=1}^N w_j(F_j (s,h))\, P(F_j (s,h)) - \sum_{j=1}^N w_j(s,h)\,P(s,h)
\label{eq:master-equation-dummy}
\end{align} 
and if the constraints $s_{ia}=s_{ib}=\ldots$ and $h_{ia}=\epsilon_a(F_is^a)-\epsilon_a(s^a))/2s_{ia}$
are satisfied initially they will be so for all time. 
Master equation (\ref{eq:master-equation-dummy}) in the dummy variables
will then describe the same dynamics as master equation (\ref{eq:master-equation}) in the original variables.
The observable of the dummy variables which multiplies the (generalized temperatures) $d(s',h')$ in (\ref{eq:auxiliary-LCS-dummy}) is
\begin{equation}
 O({s,h};s',h') = \sum_i 1_{s',\overline{s_{i}}}1_{h',\sum_a h_{ia}}
\label{eq:observable-dummy}
\end{equation} 
and the corresponding expectation value is
\begin{equation}
 \mu(s',h') = \frac{1}{N}\sum_i \<1_{s',\overline{s_{i}}}1_{h',\sum_a h_{ia}}\>
\label{eq:observable-average}
\end{equation} 
where the expectation is taken with respect to (\ref{eq:auxiliary-LCS-dummy}). 
The time derivatives in  
(\ref{eq:mu-beta-dynamics}) read for this case     
\begin{eqnarray}
 \frac{d\mu(s',h')}{dt} &=& \frac{1}{N}\sum_{j=1}^N \<w_j(\sum_a h_{ja},\overline{s_{j}}) ( O(F_j({ s,h});s',h')-O({s,h};s',h'))\> \nonumber \\
&=& \frac{1}{N}\sum_{j=1}^N \<w_j(h',s') (1_{-s',\overline{s}_{j}} - 1_{s',\overline{s_{j}}} )1_{h',\sum_a h_{ia}}\> + \nonumber \\
&&\frac{1}{N}\sum_{j=1}^N  \< w_j(\sum_a h_{ja},\overline{s_{j}}) \sum_{b\in\partial j}\sum_{i\in\partial b \setminus j} 1_{s',\overline{s_{i}}} (1_{h',F_j\sum_a h_{ia}}-1_{h',\sum_a h_{ia}})\> 
\label{eq:dynamics-Ising-dummy}   
\end{eqnarray}
The averages in the first line of the second equality above are analogous to the simpler terms previously derived for
the magnetization and energy approximation and can be written in terms of the $\mu$'s themselves:
\begin{eqnarray}\label{eq:1st-line}
 \frac{d\mu(s',h')}{dt}\Big|_{\mbox{\footnotesize{1\up{st} line}}}&=&\frac{1}{2}\left(1+s' \tanh(\beta h')\right)\mu(-s',h') - \frac{1}{2}\left(1-s'\tanh(\beta h')\right)\mu(s',h') \nonumber
\end{eqnarray}
The expression in the second line of (\ref{eq:dynamics-Ising-dummy}) is on the other hand more complicated.
We start by writing it out explicitly to be
\begin{align} \label{eq:2nd-line} \nonumber 
 \frac{d\mu(s',h')}{dt}\Big|_{\mbox{\footnotesize{2\up{nd}  line}}}=&\, \frac{1}{N}\sum_{j=1}^N  \sum_{b\in\partial j}\sum_{i\in\partial b \setminus j} \frac{1}{\Omega} \sum_{s_{jc}, h_{jc}}\sum_{s_{id}, h_{id}} 1_{s_{ia}=s_{ib}=\dots=\bar s_{i}} 1_{s_{ja}=s_{jb}=\dots=\bar s_{j}} 1 _{h_{jb},Js_{ib}} 1 _{h_{ib},Js_{jb}} \\ \nonumber
 & \mbox{e}^{d(s_{jc},\sum_{a}h_{ja})} \mbox{e}^{d(s_{id},\sum_{a}h_{ia})} \prod_{e \in j \, \backslash \, b } m_{e \to je} (s_{je}, h_{je}) \prod_{l \in i \, \backslash \, b } m_{l \to il} (s_{il}, h_{il}) \\ 
 & \frac 1 2 (1 - \overline{s_{j}}\tanh(\beta (\sum_{a} h_{ja}))) [(1_{h',F_j\sum_a h_{ia}}-1_{h',\sum_a h_{ia}})] 
\end{align}
where $\Omega$  is a normalization factor. We observe that by using the constraints $1_{s_{ia}=s_{ib}=\dots=\bar s_{i}}$ and $1_{s_{ja}=s_{jb}=\dots=\bar s_{j}}$ the sum over $\sum_{s_{jc},s_{id}}$ becomes simply $\sum_{s_j, s_i }$ and also that we can write $\sum_{a} h_{ja} = h_{jb} + \sum_{a \neq b} h_{ja}= - Js_{ib} +  \sum_{a \neq b} h_{ja}$ where the last equality is enforced by the constraint $ 1 _{h_{jb},Js_{ib}}$. An analogous relation holds for $ \sum_a h_{ia}$. Hence substituting these relations in (\ref{eq:2nd-line}) we get
\begin{align} \nonumber \label{eq:2nd-line-s1}
 \frac{d\mu(s',h')}{dt}\Big|_{\mbox{\footnotesize{2\up{nd} line}}}&= \frac{1}{N}\sum_{j=1}^N  \sum_{b\in\partial j}\sum_{i\in\partial b \setminus j} \frac{1}{\Omega} \sum_{s_{j}, h_{jc}}\sum_{s_{i}, h_{id}} \mbox{e}^{d(s_{j},Js_{i} +  \sum_{a \neq b} h_{ja})} \mbox{e}^{d(s_{i},Js_{j} +  \sum_{a \neq b} h_{ia})}\\ \nonumber
   &\prod_{e \in j \, \backslash \, b } m_{e \to je} (s_{j}, h_{je})  \prod_{l \in i \, \backslash \, b } m_{l \to il} (s_{i}, h_{il}) \frac 1 2 (1 - \overline{s_{j}}\tanh(\beta (Js_{ib} +  \sum_{a \neq b} h_{ja})))\\ 
& [1_{s',\bar s_i}(1_{h', Js_j +\sum_{a \neq b} h_{ia}}-1_{h',Js_j +\sum_{a \neq b}  h_{ia} })]
\end{align}
As we previously did in equation (\ref{eq:mess-exp}), we can write a probability distribution of two spin variables in an exponential form: $\mu_{i\to (i+1)}(s_{i},\tau_{i})\propto \exp\left(\theta_{i}J s_{i}+\nu_{i}\tau_{i}+\eta_{i}J s_{i}\tau_{i}\right)$. In the $\{s,h\}$ formalism used above this can be translated as $m_{e\to je}(s_{j},h_{je})\propto \exp(\theta J s_{j}+\nu h_{je}/J + \eta s_{j}h_{je})$ since $h_{je} = J \tau_j$ and we implicitly also assumed homogeneity ($\theta_j=\theta,\nu_j=\nu,\eta_j=\eta$) for all $j$. With this exponential representation, we can rewrite (\ref{eq:2nd-line-s1}) as:
\begin{align} \nonumber \label{}
 \frac{d\mu(s',h')}{dt}\Big|_{\mbox{\footnotesize{2\up{nd} line}}}&= \frac{1}{N}\sum_{j=1}^N  \sum_{b\in\partial j}\sum_{i\in\partial b \setminus j} \frac{1}{\Omega} \sum_{s_{j}, h_{jc}}\sum_{s_{i}, h_{id}} \mbox{e}^{d(s_{j},Js_{i} +  \sum_{a \neq b} h_{ja})} \mbox{e}^{d(s_{i},Js_{j} +  \sum_{a \neq b} h_{ia})}\\ \nonumber
 &\prod_{e \in j \, \backslash \, b } \mbox{e}^{\theta J s_{j}- \nu h_{je}/J - \eta s_{j}h_{je}}  \prod_{l \in i \, \backslash \, b } \mbox{e}^{ \theta J s_{i}-\nu h_{il}/J - \eta s_{i}h_{il}} \frac 1 2 (1 - \overline{s_{j}}\tanh(\beta (Js_{ib} +  \sum_{a \neq b} h_{ja}))) \\ \nonumber
 & [1_{s',\bar s_i}(1_{h', -Js_j +\sum_{a \neq b} h_{ia}}-1_{h',Js_j +\sum_{a \neq b}  h_{ia}})] \\ \nonumber
&= \frac{1}{N}\sum_{j=1}^N  \sum_{b\in\partial j}\sum_{i\in\partial b \setminus j} \frac{1}{\Omega} \sum_{s_{j}, h_{jc}}\sum_{s_{i}, h_{id}} \mbox{e}^{d(s_{j},Js_{i} +  \sum_{a \neq b} h_{ja})} \mbox{e}^{d(s_{i},Js_{j} +  \sum_{a \neq b} h_{ia})} \\ \nonumber
&  \mbox{e}^{\theta J (s_{i}(|\partial j | -1)+s_{i}(|\partial i | -1))-\nu (\sum_l^{'} h_{il} +\sum_e^{'} h_{je} )/J - \eta(s_{i} \sum_l^{'} h_{il}+ s_{j} \sum_e^{'} h_{je})} \\ \nonumber 
&\frac 1 2 (1 - \overline{s_{j}}\tanh(\beta (Js_{ib} +  \sum_{a \neq b} h_{ja}))) [1_{s',\bar s_i}(1_{h', -Js_j +\sum_{a \neq b} h_{ia}}-1_{h',Js_j +\sum_{a \neq b}  h_{ia}})]
\end{align}
where in the last equality we absorbed the two products into the argument of the exponent and there we defined $|\partial j |$ as the cardinality of $j$'s neighborhood and the same for $i$. We also used the compact notation $\sum_{e}^{'}=\sum_{e \in j \, \backslash \, b}$. To further simplify the notation, we now define $l_j= \sum_{e \neq b} h_{je}$ and with $n(l_i,l_j)$ we indicate the number of ways in which certain given values of $l_i, l_j$ appear in the sum. The above equation finally reads:
\begin{align}\label{eq:joint-sh} \nonumber
\frac{d\mu(s',h')}{dt}\Big|_{\mbox{\footnotesize{2\up{nd} line}}}&=  \frac{1}{N}\sum_{j=1}^N  \sum_{b\in\partial j}\sum_{i\in\partial b \setminus j} \frac{1}{\Omega} \sum_{s_{j},s_{i}} \sum_{l_{j}, l_{i}} n(l_i,l_j)\,\,\mbox{e}^{d(s_{j},Js_{i} + l_j)} \mbox{e}^{d(s_{i},Js_{j} + l_i )} \\ \nonumber 
&\mbox{e}^{\theta J s_{j}(|\partial i | -1)-\nu l_j/J - \eta s_{j}l_j} \mbox{e}^{\theta J s_{i} (|\partial j | -1)-\nu l_i/J - \eta s_{i} l_i}\\
&\frac 1 2 (1 - \overline{s_{j}}\tanh(\beta (Js_{ib} + l_j))) [1_{s',\bar s_i}(1_{h', -Js_j +l_i}-1_{h',Js_j +l_i})]
 \end{align}
and so the complete differential equation that describes the evolution in time of the joint spin-field distribution is given by
\begin{align}\label{eq:joint-sh-final} \nonumber
 \frac{d\mu(s',h')}{dt}=&\,\frac{1}{2}\left(1+s' \tanh(\beta h')\right)\mu(-s',h') - \frac{1}{2}\left(1-s'\tanh(\beta h')\right)\mu(s',h') \\ \nonumber
+& \frac{1}{N}\sum_{j=1}^N  \sum_{b\in\partial j}\sum_{i\in\partial b \setminus j} \frac{1}{\Omega} \sum_{s_{j},s_{i}} \sum_{l_{j}, l_{i}} n(l_i,l_j)\,\,\mbox{e}^{d(s_{j},Js_{i} + l_j)} \mbox{e}^{d(s_{i},Js_{j} + l_i )} \,\, \mbox{e}^{\theta J s_{j}(|\partial i | -1)-\nu l_j/J - \eta s_{j}l_j} \\
&\mbox{e}^{\theta J s_{i} (|\partial j | -1)-\nu l_i/J - \eta s_{i} l_i} \frac 1 2 (1 - \overline{s_{j}}\tanh(\beta (Js_{ib} + l_j))) [1_{s',\bar s_i}(1_{h', -Js_j +l_i}-1_{h',Js_j +l_i})]
 \end{align}
So far we only specialized to the case of pairwise interactions. In appendix \ref{appendix3} we will further assume 
that the interactions go between neighboring spins on a line \textit{i.e.} in an Ising chain, and
show that the approach derived here is then
equivalent to the micro-canonical approach summarized in Appendix~\ref{appendix2}, compare \cite{DRTinf2,Hatchett}.
 

\section{Joint spin-field distribution formalism by using a macroscopic analysis of dynamics}
\label{appendix2}
Our goal in this section is to repeat the derivation of an equation for the dynamics of the joint spin-field distribution by following the procedure introduced in \cite{DRTinf2, Hatchett}. This derivation conceptually has two parts of which the first is similar to the ansatz of a large deviation principle (compare main text),
and the second is another way to compute the averages than in Appendix~\ref{appendix1}.
The first starts by investigating the evolution in time of a general observable $\Omega( \s)$ the properties of which are fully described by the probability distribution $P_t(\Omega)= \sum_{s} p_t( s)\, \delta[\Omega-\Omega(\s)]$. One derives a Kramers-Moyal expansion for this probability which for finite times and in 
the large system limit, where only the first term survives, gives a deterministic equation
\begin{eqnarray}\label{eq:liouville}
  \frac{\mathrm{d}}{\mathrm{d}t}\Omega =
\Big\langle\sum_i w_i( \s)\Big[\Omega(F_i \s)-\Omega(\s)\Big]\Big\rangle_{\Omega ; t}
\end{eqnarray}
In Glauber dynamics $w_i( \s)= \frac 1 2 (1-\s_i \tanh(h_i(\s)))$, and the observable of interest is here the joint spin-field distribution $\Omega \equiv D(\sigma,h_{\mu},s)= \frac 1 N \sum_i \delta_ {\sigma,s_i}\delta(h_{\mu}-h_i(s))$ ($\mu(s',h')$ in the main text). 
We first work out the discrete derivative $\Omega(F_i \s)-\Omega(\s) =D(\sigma,h_\mu;F_i s)-D(\sigma,h_\mu; s)$ which, for an Ising spin chain, is:
\begin{eqnarray}\label{eq:Ddd}
  \Delta_iD(\sigma,h_\mu; s) & = & \frac{1}{N}\sum_{j=1}^N\delta_{\sigma,F_i s_j}\delta\left(h_\mu-h_j(F_i s)\right)\label{eq:DeltaD} -\frac{1}{N}\sum_{j=1}^N\delta_{\sigma,s_j}\delta\left(h_\mu-h_j(s)\right) \nonumber \\
& = & \frac{1}{N}\delta_{-\sigma,s_i}\delta\left(h_\mu-h_i(s)\right)-\frac{1}{N}\delta_{\sigma,s_i}\delta\left(h_\mu-h_i(s)\right)  \nonumber \\ 
& + & \frac{1}{N}\sum_{j \neq i}\delta_{\sigma,s_j}\delta\left(h_\mu-h_j(F_i s)\right) -\frac{1}{N}\sum_{j \neq i}\delta_{\sigma,s_j}\delta\left(h_\mu-h_j(s)\right)
\end{eqnarray}
Then we insert the above result (\ref{eq:liouville}) which, after some manipulations, gives following expression for the dynamics of the joint spin-field distribution:
\begin{eqnarray}\label{eq:Dt} 
 \frac{\partial}{\partial t}D(\sigma,h_\mu)&=&\frac{1}{2}\left (1+\sigma\tanh(\beta h_\mu)\right)D(-\sigma,h_\mu)-\frac{1}{2}\left (1-\sigma\tanh(\beta h_\mu)\right)D(\sigma,h_\mu) \nonumber \\
&+&\sum_{\tilde\sigma}\int d \tilde{ h}\frac{1}{2}(1- \tilde{\sigma} \tanh(\beta\tilde{h})) \frac{1}{N}\sum_{i}^N \bigg\{  \langle\delta_{\tilde \s,s_i}\delta (\tilde h -h_{i}(s))) \delta_{\sigma,s_{i+1}}\delta( h_\mu -h_{i+1} (F_i s))\rr_{D}  \nonumber \\
&-&\left\langle\delta_{\tilde \s,s_i}\delta (\tilde h -h_{i}(s)))\delta_{\sigma,s_{i+1}}\delta( h_\mu -h_{i+1}(s))\right\rr_{D}\bigg\}
\end{eqnarray}
where the sub-shell average $\langle \dots \rangle_D$ above is defined by 
\begin{eqnarray} \label{def:DAver}
 \left\langle f( s)\right\rangle_{D } & = &
\frac{\sum_{s} p(s) f(s)\prod_{\s,\mu}\delta \left[D(\sigma,h_\mu) - D(\sigma,h_\mu; s)\right]}{\sum_{\hat{ s} }p(\hat s)\prod_{\sigma, \mu}\delta \left[D(\sigma,h_\mu) - D(\sigma,h_\mu; \hat s )\right]},
\end{eqnarray}
At this point the equations for the time developments of the averages, equations (\ref{eq:Dt}) or equivalently, (\ref{eq:liouville}), are exact but not closed because they depend explicitly on $p(s)$. In order to close them we
assume that the microscopic probability depends on the state of the system only through the observable of interest, in this case $p(s)= p(D(\sigma, h_{\mu};s))$. 
This assumption, called the \textit{equipartition assumption} in \cite{DRTinf1,DRTinf2,coolen2005theory}, is
analogous to the large deviation assumption used in the main text of the paper, but is stronger.
If it can be used then (\ref{def:DAver}) simplifies to
\begin{eqnarray} \label{def:DAver_simply}
 \left\langle f( s)\right\rangle_{D } & = &
\frac{\sum_{s} f(s)\prod_{\s,\mu}\delta \left[D(\sigma,h_\mu) - D(\sigma,h_\mu; s)\right]}{\sum_{\hat{ s} }\prod_{\sigma, \mu}\delta \left[D(\sigma,h_\mu) - D(\sigma,h_\mu; \hat s )\right]},
\end{eqnarray}
and by a Legendre transform we get the large deviation form
\be
p_d(s) =\frac{1}{Z_d} \exp\Big[N\sum_\sigma \sum_{\mu}\, d(\sigma,h_{\mu})D(\sigma,h_{\mu};s)\Big]\label{eq:Pd} = \frac{1}{Z_d} \exp\Big[\sum_{i=1}^N d(s_i,h_i(s))\Big],
\ee
where 
\be\label{def:Zd}
Z_d = \sum_s \exp\Big[\sum_{i=1}^N d(s_i,h_i(s))\Big]
\ee
The equipartition assumption is stronger than the assumption of a large deviation principle both
because the large deviation principle explicitly admits sub-leading terms, and because the inverse Legendre transform does
not have to be uniquely defined.

From now on we could introduce dummy variables and proceed to evaluate the marginal probabilities as in Appendix~\ref{appendix1}.
The alternative route instead proceeds directly and starts from the 
observation that the interesting physics and non-trivial part of (\ref{eq:Dt}) is contained in the terms between angular brackets. To simplify the notation and compute those terms in a more compact manner we hence define the following kernel:
\begin{eqnarray}\label{eq:A-average}\nonumber 
A[\s, h, \tilde \s, \tilde h | \hat F]&=&\Big\langle\frac{1}{2N}\sum_{i}\delta_{{\tilde \s},s_i}\, \delta (\tilde h - h_i (s)) \, \delta_{\s,s_{i+1}} \, \delta ( h\! -\! h_{i+1} (s)\!+\! 2J \hat F)\Big\rangle_{D;t} \label{} \\
&=&\frac{1}{2}\Big\langle\delta_{{\tilde \s},s_0}\, \delta (\tilde h - h_0 (s)) \, \delta_{\s,s_{1}} \, \delta ( h\! -\! h_{1} (s)\!+\! 2J \hat F)\Big\rangle_{D;t} \label{def:A}
\end{eqnarray}
where $\hat F = 0$ if there is no spin flip, and $\hat F = 1$ if there is. To arrive at (\ref{eq:A-average})
we have assumed spatial homogeneity. Equation (\ref{eq:Dt}) can then be written as
\begin{eqnarray}\label{eq:diff_Dsh} 
 \frac{\partial}{\partial t}D(\sigma,h)&=&\frac{1}{2}\left (1+\sigma\tanh(\beta h)\right)D(-\sigma,h)-\frac{1}{2}\left (1-\sigma\tanh(\beta h)\right)D(\sigma,h)\nonumber\\
&+&\sum_{\tilde\sigma}\int d \tilde{ h}\frac{1}{2}(1- \tilde{\sigma} \tanh(\beta\tilde{h})) \Big \{ A[\s, h, \tilde \s, \tilde h | 1] - A[\s, h, \tilde \s, \tilde h | 0] \Big\}
\end{eqnarray}
The averages contained in the function $A[\s, h, \tilde \s, \tilde h | \hat F]$ can be computed from the cavity method using cavities
containing more than one variable. First we rewrite the partition function (\ref{def:Zd}) as:
\bearr \label{eq:Zd} \nonumber
Z_d  &=& \sum_{\s, \tilde \s} \int dh \int d\tilde h \sum_s \, \mbox{e}^{ \, \sum_{i=1}^N d(s_i,h_i(s))} \ \delta_{{\tilde \sigma},s_i}\, \delta (\tilde h - h_i (s)) \, \delta_{\s,s_{i+1}} \, \delta ( h\! -\! h_{i+1} (F_i  s)) \\
&\equiv&  \sum_{\s, \tilde \s} \int dh \int d\tilde h \, Z_d[\s, h, \tilde \s, \tilde h| \hat F]
\eearr
where we implicitly defined also the ``marginal'' partition function $Z_d[\s, h, \tilde \s, \tilde h| \hat F]$. 
For the chain topology we can compare (\ref{eq:A-average}) and (\ref{eq:Zd}) to write
\begin{equation}\label{eq:A-definition}
A[\s,\tilde \s, h, \tilde h|\hat F]= \frac{Z_d[\s, h, \tilde \s, \tilde h| \hat F]}{\sum_{\s, \tilde \s} \int dh \int d\tilde h \sum_s \, Z_d[\s, h, \tilde \s, \tilde h| \hat F]}
\end{equation}
Applying then the cavity method we have
\begin{align} \label{eq:Z-marginal} \nonumber
Z_d[\s, h, \tilde \s, \tilde h| \hat F]  &= \mbox{e}^{ d(\tilde \s,\tilde h)+d(\s,h+2J\s \hat F)}\sum_s \mbox{e}^{ \, \sum_{j\neq(i,i+1)}^N d(s_i,h_i(s))} \ \delta_{{\tilde \s},s_i}\, \delta (\tilde h - h_i ( s)) \, \delta_{\s,s_{i+1}} \, \delta (h\! -\! h_{i+1} (F_i  s)) \\ 
&= \mbox{e}^{ d(\tilde \s,\tilde h)+d(\s,h+2J\s \hat F)} \sum_{s_{-1},s_{0},s_{1},s_{2}} \Q_{-1}(s_{-1},J s_0) \Q_{2}(s_{2},J s_{1}).
\end{align}
where $\Q_{-1}$ and $\Q_{2}$ are two functions defined recursively as  $\Q_{-1}(s_{-1},J s_0)=\Q_{i-1}(s_{i-1}, s_i J)= \sum_{s_{i-2}} \exp{\{d(s_{i-1}, h_{i-1}(s) )\}}\, \Q_{i-2}(s_{i-2}, s_{i-1} J)$.
Once the cavity assumption is made, the chain separates into two independent branches and hence the sum over $s$ can be factorized and rewritten by using 
the two functions $\Q_{-1}$ and $\Q_{2}$, each corresponding to one branch of the chain. 
Since these are functions of two Boolean variables they can be written in an exponential form as $\Q_{i-1}(s_{i-1}, s_i J) \propto \exp{\{ \nu s_{i-1} + \theta s_i+ \eta s_{i-1} s_i \}}$, where again, because of homogeneity $\nu_i= \nu, \theta_i=\theta, \eta_i=\eta$. 
Hence substituting this exponential form for $\Q$ in (\ref{eq:Z-marginal}) we can hence explicitly compute (\ref{eq:A-definition}) 
\begin{eqnarray} \nonumber
A[\s, h, \tilde \s, \tilde h | \hat F] &=& \frac{1}{Z_d} \sum_{s_{-1},s_{0}, s_{1},s_{2}} \mbox{e}^{d(s_0,\, h_0(s))} \mbox{e}^{d(s_1,\, h_1(s))} \mbox{e}^{\nu(s_{-1}+s_2)+\theta J(s_{0}+s_1)+\eta J(s_{-1}s_0+s_1 s_2)} \\ 
&& \delta_{{\tilde \s},s_0}\, \delta (\tilde h - h_0 ( s)) \, \delta_{\s,s_{1}} \, \delta( h\! -\! h_{1} (\hat F s))
\end{eqnarray}
where we have defined $Z_d = \sum_{\s, \tilde \s} \int dh \int d\tilde h \sum_s \, Z_d[\s, h, \tilde \s, \tilde h| \hat F]$ and observed that has the same expression both with $\hat F = 1$ that $\hat F=0$. This expression can now be substituted back into (\ref{eq:diff_Dsh}) in order to obtain the explicit expression for the joint spin-field distribution dynamics. After summing over $\tilde \sigma$ and integrating over $\tilde h$, the final results reads:
\begin{align}\label{eq:diff_Dsh_final}\nonumber 
 \frac{\partial}{\partial t}D(\sigma,h)&= \frac{1}{2}\left (1+\sigma\tanh(\beta h)\right)D(-\sigma,h)-\frac{1}{2}\left (1-\sigma\tanh(\beta h)\right)D(\sigma,h)\\ \nonumber
&+  \frac{1}{Z_d}\sum_{s_{-1},s_{0}, s_{1},s_{2}} \frac{1}{2} (1- s_0 \tanh(\beta \, h_0(s)))\, \mbox{e}^{d(s_0,\, h_0(s))} \mbox{e}^{d(s_1,\, h_1(s))}  \\ 
& \mbox{e}^{\nu(s_{-1}+s_2)+\theta J(s_{0}+s_1)+\eta J(s_{-1}s_0+s_1 s_2)} \delta_{\s,s_{1}} \big(\, \delta( h\! -\! h_{1} (\hat F s))- \delta( h\! -\! h_{1} (\hat s))\big)
\end{align}
To conclude we observe that the equation $\Q_{i-1}(s_{i-1}, s_i J)= \sum_{s_{i-2}} \exp{\{d(s_{i-1}, h_{i-1}(s) )\}} \Q_{i-2}(s_{i-2}, s_{i-1} J)$ seen above is basically equivalent to (\ref{eq:mess-exp}) and therefore the parameters $\nu, \theta, \eta$ introduced here satisfy the same equations shown in (\ref{eq:cavity-field}).

\section{Equivalence between the large deviation approach and the macroscopic analysis of dynamics}\label{appendix3}
In this section we want to show that, for an Ising spin chain with nearest neighbors interactions, the two different formalisms developed in the paper for the joint spin-field distribution dynamics (Appendix \ref{appendix1} and \ref{appendix2}) are equivalent. Recall the second term on the right hand side of the equation (\ref{eq:joint-sh}) obtained in Appendix~(\ref{appendix1})
\begin{align} \nonumber
 \frac{d\mu(s',h')}{dt}\Big|_{\mbox{\footnotesize{2\up{nd}  line}}}=& \frac{1}{N}\sum_{j=1}^N  \sum_{b\in\partial j}\sum_{i\in\partial b \setminus j} \frac{1}{\Omega} \sum_{s_{j},s_{i}} \sum_{l_{j}, l_{i}} n(l_i,l_j)\,\,\mbox{e}^{d(s_{j},Js_{i} + l_j)} \mbox{e}^{d(s_{i},Js_{j} + l_i )} \\ \nonumber
 &\mbox{e}^{a J s_{j}(|\partial i | -1)-b l_j/J - cs_{j}l_j} \, \mbox{e}^{a J s_{i} (|\partial j | -1)-b l_i/J - cs_{i} l_i} \\
 &\frac 1 2 (1 - s_{j}\tanh(\beta (Js_{ib} + l_j))) [1_{s',\bar s_i}(1_{h', -Js_j +l_i}-1_{h',Js_j +l_i})]
 \end{align}
We now specify this general expression to a chain.
The sum $\frac{1}{N}\sum_{j=1}^N  \sum_{b\in\partial j}\sum_{i\in\partial b \setminus j}$ is then simply equal to one
and since every node has precisely two neighbours $|\partial j|=|\partial i|=2$. 
By considering $i=j+1$ as one of the $j$'s neighbours, the two variables ($l_j,l_i$) introduced in (\ref{appendix1}) become simply $l_j = \sum_{e \neq b} h_{jb}= Js_{j-1}$
and $l_i=l_{j+1}= J s_{j+2}$; the multiplicity is hence $n(l_j,l_i)=1$. We also observe than summing over ($l_j,l_i$) corresponds to summing over ($s_{j-1}, s_{j+2}$) and so by making these substitutions the equation simplifies to
\begin{align}\label{eq:final-joint-sh2} \nonumber 
\frac{d\mu(s',h')}{dt}\Big|_{\mbox{\footnotesize{2\up{nd}  line}}}&= \frac{1}{\Omega} \sum_{s_{j-1}, s_{j}}  \sum_{s_{j+1}, s_{j+2}} \mbox{e}^{d(s_{j},Js_{j+1} + Js_{j-1})} \mbox{e}^{d(s_{j+1},Js_{j} + Js_{s+2} )}  \\ \nonumber
&\frac 1 2 (1 - s_{j}\tanh(\beta \, J(s_{j+1} + s_{j-1})) \mbox{e}^{\theta J (s_{j}+s_{j+1})+\nu (s_{j-1}+s_{j+2}) +\eta J (s_{j}s_{j-1} + s_{j+1}s_{j+2})} \\  
&[1_{s',\overline{s_j}}(1_{h', -Js_j +J s_{j+2}}-1_{h',J(s_j + s_{j+2})})] 
\end{align}
\noindent
If we relabel the spin variables as $s_{j-1}=s_{-1}$, $s_{j}=s_{0}$, $s_{j+1}=s_{1}$ and $s_{j+2}=s_{2}$, the complete differential equation for the joint spin-field distribution looks like:
 \begin{align}\label{eq:final-joint-sh} \nonumber 
\frac{d\mu(s',h')}{dt}=&\frac{1}{2}\left(1+s'\tanh(\beta h')\right)\mu(-s',h') - \frac{1}{2}\left(1-s'\tanh(\beta h')\right)\mu(s',h') \\ \nonumber
&+\frac{1}{\Omega} \sum_{s_{0},s_{1}} \sum_{s_{-1}, s_{2}}\,\, \mbox{e}^{d(s_{0},h_{0}(s))}\,\,  \mbox{e}^{d(s_{1}, h_{1}(s) )}  \,\, \mbox{e}^{\theta J (s_{0}+s_{1})+\nu (s_{-1}+s_{2}) + \eta J (s_{0}s_{-1} + s_{1}s_{2})} \\ 
&\frac 1 2 (1 - s_{0}\tanh(\beta \,h_0(s)))[1_{s',s_1}(1_{h', -Js_0 +J s_{2}}-1_{h',J(s_0 + s_{2})})] 
 \end{align}
which is equivalent to equation (\ref{eq:diff_Dsh_final}) if we finally rename $\mu(s',h')$ to read $D(\sigma, h)$, and $\Omega$ to read $Z_d$.


\section{The Amari hierarchy}\label{appendix4}
This appendix is intended as a pointer to the literature on Information Geometry where
closely similar concepts have been developed some time ago, see~\cite{Amari1987} and especially~\cite{Amari2001}. 
Begin by considering the space $E$ of all probability distributions on $N$ spins which can be written in exponential
form as
\begin{equation}
\log p = \sum_i \theta_i\sigma_i + \sum_{ij}\theta_{ij}\sigma_i\sigma_j+\ldots-\psi(\theta)
\label{eq:general-Boolean}
\end{equation}
and consider a partition of the interactions terms in increasing sets $S_0\subset S_1\subset\ldots\subset S_L\subset S$
where $S_0$ is the empty set and $S$ contains all the interactions. One natural partition is to take
$S_1$ all terms depending one spin (all $\theta_i$'s), $S_2$ all terms depending on one or two spins (all $\theta_i$'s and $\theta_{ij}$'s), 
and so on. Here we will assume a partition which follows the terms of the assumed
auxiliary Gibbs distribution (\ref{eq:auxiliary-Gibbs}). We then consider the foliation 
$E_0\subset E_L\subset E$
where $E_0$ has only one element, the uniform measure, and  $E_L$ is the subfamily of 
distributions taking non-zero coefficients only in the set $S_L$. 
Referring to~\cite{Amari1987} for background on Information Geometry we state that
this foliation is an $e$-flat hierarchical structure as defined in~\cite{Amari2001}.
Furthermore, the $m$-projection (see~\cite{Amari1987,Amari2001}) of an element $p\in E$ on $E_L$ is denoted $p^{(L)}$ and is,
in the case at hand,
defined as the probability distribution having the same expectation values as $p$ for all interactions in $S_L$, and 
all interaction coefficients beyond $S_L$ zero. 
It follows from these definitions that
the reduced dynamics described by (\ref{eq:mu-beta-dynamics}) 
is the $m$-projection of (\ref{eq:master-equation}) on the submanifold $E_L$.
We note that we assume the full probability distribution $P$ always to be close to the submanifold
so that the projection is only of the infinitesimal increment of $P$ to $P'=P+\delta P$.
Amari in~\cite{Amari2001} also constructs a dual foliation which we write 
$M\subset M_L\subset M_0$ 
where $M$ is the uniform distribution, $M_0$ contains all the probability distributions parametrized as a mixture model, 
and $M_L$, where the expectation values of all interaction terms in $S_L$ are zero but the
other can take any value~\footnote{Note that the explicit enumeration of the leaves of the foliation
of the $M$-flat structure in eq.~56 on page~1706 of~\protect\cite{Amari2001} is the opposite,
we here follow the description around eq.~36 on page~1704.}.
This foliation is dual to 
$E_0\subset E_L\subset E$ in the sense that
any probability distribution can be parametrized by 
combining a coordinate in $E_L$ (parameters of the exponential family in $S_L$)
and a coordinate in $M_L$ (zero expectation values for the terms in $S_L$, free values
of expectation values beyond $S_L$). This combination, called the $k$-cut mixed
coordinate system in \cite{Amari2001}, shows that $M_L$ and $E_L$ are orthogonal and complementary at every point,
and the error we make in the dimensional reduction (projection on $E_L$) is hence the projection of the 
probability increment $\delta P$ on $M_L$.

What this means is simply that any probability distribution can be parametrized both as
an exponential family (\ref{eq:general-Boolean}) ($e$-coordinates) and as a mixture model
($m$-coordinates). The submanifold $P^{\hbox{aux}}$ has a simple description in the
$e$-coordinates, but is a (perhaps complicated) hypersurface in the $m$-coordinates
as well as in the $k$-cut mixed coordinates. If we change a point on $P^{\hbox{aux}}$ from
$P$ to $P'=P+\delta P$ then this leads to a change in its $k$-cut mixed coordinates
where the first part (the generalized temperatures) change as (\ref{eq:beta-dynamics})
while the second part (the expectation values) changes as the first term on the right hand
side of (\ref{eq:error-Q}). At the same time the projection of $P'=P+\delta P$ on
 $P^{\hbox{aux}}$ changes in the first part of its $k$-cut mixed coordinates in the same
way as $P$ itself, while the second part changes as the second term on the right hand
side of (\ref{eq:error-Q}). The difference between these two quantities hence gives
how much  $P'=P+\delta P$ differs from its projection in directions orthogonal to $P^{\hbox{aux}}$.
 
\bibliographystyle{unsrt}
\bibliography{mozeika,general}
\end{document}